\documentclass[11pt]{article}

\usepackage[T1]{fontenc}
\usepackage{jheppub}
\usepackage{wasysym}

\usepackage{tabularx}
\usepackage{physics}
\usepackage{tikz}

\usepackage{graphicx} 
\usepackage{ae}

\usepackage{amsfonts,amsmath,amsthm,amssymb,amsbsy}
\usepackage{mathtools}

\newcommand{\vertsGreen}[1]{%
    \mathcal{V}^{+}_{#1}
}

\newcommand{\vertsRed}[1]{%
    \mathcal{V}^{-}_{#1}
}

\title{Positive and Negative Ladders in Loop Space}

\author[]{Ross Glew and}\emailAdd{r.glew@herts.ac.uk}
\author[]{Tomasz \L ukowski}\emailAdd{t.lukowski@herts.ac.uk}

\affiliation[]{Department of Physics, Astronomy and Mathematics, \\ University of Hertfordshire, \\  Hatfield, Hertfordshire, AL10 9AB, United Kingdom}

\abstract{
Motivated by a new term-wise factorised formula for the two-loop MHV integrand for scattering amplitudes in $\mathcal{N}=4$ super Yang-Mills (SYM), together with recent results for the five-point negative ladders in loop space, we present the canonical forms for general ladders in loop space for an arbitrary number of particles to all loops. We make use of the graphical notation introduced in the negative geometries literature, where each loop momentum is represented as a vertex, and mutual positivity (resp. negativity) conditions as a positive (resp. negative) edge. In this paper we extend this notation to include the notion of chambers of the one-loop momentum amplituhedron. Equipped with this new graphical notation, we find the canonical form of the $L$-loop (negative/positive) ladders for all MHV$_n$ amplitudes. Our final formula is remarkably simple and reminiscent of the chiral pentagon expansion of the one and two loop momentum amplituhedron. It expresses ladder contributions as sums over maximal cuts, with each term appearing in the sum factorising into products of either chiral pentagons or their simple generalisations. }

\begin{document}

\maketitle
\pagebreak
\section{Introduction}
In recent years much progress has been made towards the reformulation of scattering amplitudes in terms of positive geometries \cite{Arkani-Hamed:2017tmz}. Within this framework scattering amplitudes are expressed as the differential canonical form of a particular positive geometry. The prime example of this approach is the representation of the all-loop integrand of planar $\mathcal{N}=4$ SYM as the canonical form of the amplituhedron \cite{Arkani-Hamed:2013jha}. Since the discovery of the amplituhedron, scattering amplitudes in other theories such as ABJM theory \cite{He:2022cup,He:2023rou} and $\tr \left(\phi^3\right)$ theory \cite{Arkani-Hamed:2017mur}, as well as other physical quantities such as correlations function \cite{Eden:2017fow,He:2024xed}, have also been described using positive geometries.

Originally proposed in momentum twistor space, the analogue of the amplituhedron has since been studied in spinor-helicity space, where the corresponding positive geometry is referred to as the momentum amplituhedron \cite{Damgaard:2019ztj,Ferro:2022abq}. More recently, it has been translated directly into the space of dual momenta with split signature $\mathbb{R}^{2,2}$ \cite{Ferro:2023qdp}, with similar results previously obtained for the ABJM theory in the three-dimensional Minkowski space in \cite{Lukowski:2023nnf}. It is the dual-momentum space perspective which we adopt in this paper where we focus on MHV$_n$ integrands. In dual-momentum space the scattering data upon which the amplituhedron is defined consists of a null polygon ${\bf x} = \{ x_1, \ldots, x_n\}$, whose edges define the momenta of the scattering particles, together with $L$ additional points $\left( y_1,\ldots,y_L\right)$ associated to the loop momenta. At tree-level the amplituhedron places constraints upon the configuration of the null polygon ${\bf x}$, in particular the vertices of the null polygon are constrained to be positively separated from one another. Viewing the tree-level data as fixed, the $L$-loop amplituhedron is then defined as the collection of points $(y_1,\ldots,y_L)$ positively separated from each vertex of the null polygon and additionally constrained to be positively separated from one another, that is they satisfy the constraints\footnote{Here we have omitted the additional sign flip conditions needed to define the amplituhedron but will return to the full definition in the main text.} 
\begin{align}
\forall \ i \in [n] \text{ and } \forall  \ a,b \in [L] \text{ we have }(y_a-x_i)^2 > 0 \text{ and } (y_a-y_b)^2 ,
\label{eq:amp_loop_intro}
\end{align}
where for any positive integer $p$ we define $[p]=\{1,2,\ldots,p\}$. The first set of conditions constrains each loop momentum to be in the one-loop amplituhedron region, which we refer to as the one-loop fiber $\Delta({\bf x})$, and the second set of conditions impose mutual positivity amongst loop momenta. 

The structure of the one-loop fiber $\Delta({\bf x})$, most importantly its vertex set, has been studied for arbitrary helicity in \cite{Ferro:2023qdp}. In this paper we will be interested in the MHV results, in which case, independent of the tree-level configuration, the vertex set of $\Delta({\bf x})$ consists of the vertices ${\bf x}$ of the null polygon, together with a set of quadruple cut points $\ell^*_{ij}$ for $1<i<j<n$ defined to satisfy the quadruple cut constraints
\begin{equation}
(\ell^*_{ij}-x_i)^2=(\ell^*_{ij}-x_{i+1})^2=(\ell^*_{ij}-x_j)^2=(\ell^*_{ij}-x_{j+1})^2=0\,.
\end{equation}
From the positivity conditions \eqref{eq:amp_loop_intro}, all points inside the fiber $\Delta(\bf x)$ are positively separated from all $x_i$, however the distances to the quadruple cuts $\ell_{ij}^*$ generally vary inside $\Delta(\bf x)$. It is therefore natural to consider how the signs of distances to the quadruple cut vertices $\ell^*_{ij}$ decompose the one-loop region into smaller regions. Using this decomposition we can associate to each point $y \in \Delta({ \bf x})$ a sign pattern for the distances $(y-\ell^*_{ij})^2$ and define the {\it one-loop chambers}, $\mathfrak{C} = \{\mathfrak{c}_{1},\ldots,\mathfrak{c}_{[n_{\mathfrak{C}}]}\}$, to be subsets of $\Delta({\bf x})$ with a fixed sign pattern. The question of how the one-loop fiber decomposes into one-loop chambers was recently answered in \cite{Ferro:2024vwn}. Utilising a connection to the $m=2$ amplituhedron, and ultimately the hypersimplex \cite{Parisi:2021oql}, it was found that the one-loop fiber decomposes into $n_{\mathfrak{C}}=E_{3,n-1}$ many one-loop chambers, where $E_{3,n-1}$ are the Eulerian numbers. In this paper we will introduce the following graphical notation for the chamber decomposition of the one-loop amplituhedron
\begin{align}
\begin{tikzpicture}[baseline=-0.5ex,scale=0.8]
        \draw[ultra thick,green] (0,0) ;
         \node[below=4pt] at (0,0) {$y$};
        \node[fill=black, circle, minimum size=5pt, inner sep=0pt, line width=1pt, draw=black] (A) at (0,0) {};
\end{tikzpicture} & =\sum_{\alpha \in [n_{\mathfrak{C}}]} \begin{tikzpicture}[baseline=-0.5ex,scale=0.8]
        \draw[ultra thick,green]  (-1,0.5)--(-2,0);
        \draw[ultra thick,red]  (-1,-0.5)--(-2,0);
        \node[right=4pt] at (-1,0.5) {$\vertsGreen{\alpha}$};
        \node[right=4pt] at (-1,-0.5) {$\vertsRed{\alpha}$};
         \node[below=4pt] at (-2,0) {$y$};
         \node[rectangle, draw, fill=black, minimum width=5 pt, minimum height=5 pt,inner sep=0pt] (pt1) at (-1,0.5) {};
         \node[rectangle, draw, fill=black, minimum width=5 pt, minimum height=5 pt,inner sep=0pt] (pt2) at (-1,-0.5) {};
        \node[fill=black, circle, minimum size=5pt, inner sep=0pt, line width=1pt, draw=black] at (-2,0) {};
\end{tikzpicture},
\label{eq:intro_decomp}
\end{align}
where on the left hand side we have the canonical form for the one-loop amplituhedron, and the term appearing in the sum on the right hand side is the canonical form for the one-loop chamber $\mathfrak{c}_\alpha$. The sets $\vertsGreen{\alpha}$ and $\vertsRed{\alpha}$ contain quadruple cuts $\ell_{ij}^*$ for which the distances $(y-\ell_{ij}^*)^2$ are negative (red) or positive (green) for all points $y$ in a given chamber. The utility of the decomposition \eqref{eq:intro_decomp}  becomes most apparent at two loops where in \cite{Ferro:2024vwn} it allowed for a re-writing of the two-loop MHV integrand in the following (term-wise) factorised form 
\begin{align}
\begin{tikzpicture}[baseline=-0.5ex,scale=0.8]
        \draw[ultra thick,green] (0,0) -- (1,0);
         \node[below=4pt] at (0,0) {$y_1$};
         \node[below=4pt] at (1,0) {$y_2$};
        \node[fill=black, circle, minimum size=5pt, inner sep=0pt, line width=1pt, draw=black] (A) at (0,0) {};
        \node[fill=black, circle, minimum size=5pt, inner sep=0pt, line width=1pt, draw=black] (B) at (1,0) {};
\end{tikzpicture} & = \sum_{1 \leq i \leq j \leq n} \begin{tikzpicture}[baseline=-0.5ex,scale=0.8]
        \draw[ultra thick,green] (0,0) -- (1,0);
         \node[below=4pt] at (0,0) {$y_1$};
         \node[below=4pt] at (1,0) {$\ell^*_{ij}$};
        \node[fill=black, circle, minimum size=5pt, inner sep=0pt, line width=1pt, draw=black] (A) at (0,0) {};
        \node[rectangle, draw, fill=black, minimum width=5 pt, minimum height=5 pt,inner sep=0pt] (B) at (1,0) {};
\end{tikzpicture} \wedge \begin{tikzpicture}[baseline=-0.5ex,scale=0.8]
        \draw[ultra thick,black] (-1,0) -- (0,0) -- (1,0);
        \node[below=4pt] at (-1,0) {$\ell^*_{ij}$};
         \node[below=4pt] at (0,0) {$y_1$};
         \node[below=4pt] at (1,0) {$y_2$};
          \node[rectangle, draw, fill=black, minimum width=5 pt, minimum height=5 pt,inner sep=0pt] (pt1) at (-1,0) {};
        \node[rectangle, draw, fill=black, minimum width=5 pt, minimum height=5 pt,inner sep=0pt] (A) at (0,0) {};
        \node[fill=black, circle, minimum size=5pt, inner sep=0pt, line width=1pt, draw=black] (B) at (1,0) {};
\end{tikzpicture}.
\label{eq:two_loop_chiral_box_intro}
\end{align}
On the left hand side we have the two-loop amplituhedron form, and on the right hand side the first factor is a form in $y_1$ only, whereas the second factor is a form in $y_2$ only with $y_1$ treated as a fixed point, given by the familiar chiral box integrands. This term-wise factorisation property of the above formula was referred to as a fibration in \cite{Ferro:2024vwn}.

An interesting question to consider is how the fibration formula \eqref{eq:two_loop_chiral_box_intro} might generalise to higher loops. Whereas this is a difficult question for the amplituhedron in general, a simpler target is to consider the geometries obtained by relaxing, or flipping the sign of, the mutual positivity conditions $(y_a-y_b)^2$ amongst loops. To encode the set of mutual positivity constraints we follow the graphical notation introduced in \cite{Arkani-Hamed:2021iya} where each loop variable is represented as a vertex and mutual positivity or negativity between loop variables $y_a$ and $y_b$ is represented by a green or red link respectively. Following the convention introduced in \cite{Brown:2023mqi} we refer to the graphs constructed in this way as {\it graphs in loop space}. A particularly simple family of graphs in loop space are given by the ladder topologies with all negative links
\begin{align}
\begin{tikzpicture}[baseline=-0.5ex,scale=0.7]
        \draw[ultra thick,red] (0,0) -- (1,0)--(1.2,0);
        \draw[ultra thick,red] (1.8,0) -- (2,0) -- (3,0);
        \node[fill=black, circle, minimum size=4pt, inner sep=0pt, line width=1pt, draw=black] (A) at (0,0) {};
        \node[fill=black, circle, minimum size=4pt, inner sep=0pt, line width=1pt, draw=black] (B) at (1,0) {};
        \node[fill=black, circle, minimum size=4pt, inner sep=0pt, line width=1pt, draw=black] (C) at (2,0) {};
        \node[fill=black, circle, minimum size=4pt, inner sep=0pt, line width=1pt, draw=black] (D) at (3,0) {};
        \node[] at (1.55,0) {{\color{red} \small \ldots}};
         \node[below=4pt] at (0,0) {$y_1$};
         \node[below=4pt] at (1,0) {$y_2$};
         \node[below=4pt] at (3,0) {$y_{L+1}$};
\end{tikzpicture}  = \Omega_{y_1\ldots y_{L+1}}\left[\{ y_1,\ldots,y_{L+1} \in \Delta({\bf x}) : (y_{a+1}-y_a)^2<0 \text{ for all } a \in [L] \}\right]. \notag 
\end{align}
Such graphs with all negative links have been studied under the name of {\it negative geometries}, and all loop formulae have already been obtained for the four and five-point ladders in \cite{Arkani-Hamed:2021iya,Chicherin:2024hes}. The main result of this paper will be to generalise these results to an arbitrary number of particles, for all loop orders. In particular, we find the idea of fibration extends to all ladder geometries where the solution takes the following graphical form
\begin{align}
 \begin{tikzpicture}[baseline=-0.5ex,scale=0.7]
        \draw[ultra thick,red] (0,0) -- (1,0)--(1.2,0);
        \draw[ultra thick,red] (1.8,0) -- (2,0) -- (3,0);
        \node[fill=black, circle, minimum size=4pt, inner sep=0pt, line width=1pt, draw=black] (A) at (0,0) {};
        \node[fill=black, circle, minimum size=4pt, inner sep=0pt, line width=1pt, draw=black] (B) at (1,0) {};
        \node[fill=black, circle, minimum size=4pt, inner sep=0pt, line width=1pt, draw=black] (C) at (2,0) {};
        \node[fill=black, circle, minimum size=4pt, inner sep=0pt, line width=1pt, draw=black] (D) at (3,0) {};
        \node[] at (1.55,0) {\textcolor{red}{{\small \ldots}}};
         \node[below=4pt] at (0,0) {$y_1$};
         \node[below=4pt] at (1,0) {$y_2$};
         \node[below=4pt] at (3,0) {$y_{L+1}$};
\end{tikzpicture}=\sum_{\ell^*_{kl} \in \mathcal{V}^- } \sum_{\ell^*_{ij} \in \mathcal{V}^- }\begin{tikzpicture}[baseline=-0.5ex,scale=0.6]
        \draw[ultra thick,red] (0,0) -- (1,0);
         \node[below=4pt] at (0,0) {$y_1$};
         \node[below=4pt] at (1,0) {$\ell^*_{kl}$};
        \node[fill=black, circle, minimum size=5pt, inner sep=0pt, line width=1pt, draw=black] (A) at (0,0) {};
        \node[rectangle, draw, fill=black, minimum width=5 pt, minimum height=5 pt,inner sep=0pt] (B) at (1,0) {};
\end{tikzpicture} \wedge  \begin{tikzpicture}[baseline=-0.5ex,scale=0.7]
        \draw[ultra thick,black] (-1,0) -- (1,0);
        \draw[ultra thick,red] (1,0)--(1.2,0);
        \draw[ultra thick,red] (1.8,0) -- (2,0) -- (3,0);
        \node[rectangle, draw, fill=black, minimum width=5 pt, minimum height=5 pt,inner sep=0pt] at (-1,0) {};
         \node[rectangle, draw, fill=black, minimum width=5 pt, minimum height=5 pt,inner sep=0pt] at (0,0) {};
        \node[fill=black, circle, minimum size=4pt, inner sep=0pt, line width=1pt, draw=black] at (1,0) {};
        \node[fill=black, circle, minimum size=4pt, inner sep=0pt, line width=1pt, draw=black] at (2,0) {};
        \node[rectangle, draw, fill=black, minimum width=5 pt, minimum height=5 pt,inner sep=0pt] at (3,0) {};
        \node[] at (1.55,0) {\textcolor{red}{{\small \ldots}}};
         \node[below=4pt] at (-1,0) {$\ell^*_{kl}$};
         \node[below=4pt] at (0,0) {$y_1$};
         \node[below=4pt] at (1,0) {$y_{2}$};
         \node[below=4pt] at (2,0) {$y_{L}$};
         \node[below=4pt] at (3,0) {$\ell^*_{ij}$};
\end{tikzpicture}   \wedge \begin{tikzpicture}[baseline=-0.5ex,scale=0.7]
        \draw[ultra thick,black] (-1,0) -- (1,0);
        \node[below=4pt] at (-1,0) {$\ell^*_{ij}$};
         \node[below=4pt] at (0,0) {$y_{L}$};
         \node[below=4pt] at (1,0) {$y_{L+1}$};
          \node[rectangle, draw, fill=black, minimum width=5 pt, minimum height=5 pt,inner sep=0pt] (pt1) at (-1,0) {};
        \node[rectangle, draw, fill=black, minimum width=5 pt, minimum height=5 pt,inner sep=0pt] (A) at (0,0) {};
        \node[fill=black, circle, minimum size=5pt, inner sep=0pt, line width=1pt, draw=black] (B) at (1,0) {};
\end{tikzpicture},
\end{align}
where the right hand side is written in a factorised form originating from the fibration of fibration idea \cite{Ferro:2024vwn}, and the sums are over all quadruple cuts $\ell_{ij}^*$ in the fiber geometry $\Delta(\bf x)$.  Only one of the terms appearing on the right is new compared to the two-loop result and we provide a formula for this factor in terms of simple generalisations of chiral pentagons. In the main text we will also provide similar formulae that are valid for ladders with any number of negative links replaced by positive ones.   

It is important to note that these ladder geometries not only provide a simplified version of the momentum amplituhedron, but that they are also relevant for infrared finite quantities in $\mathcal{N}=4$ SYM. In particular, in the case of the negative geometries, upon integrating all but one loop momentum variable, the integrated result can be interpreted as the expectation value of a null polygonal Wilson loop with a Lagrangian insertion \cite{Arkani-Hamed:2021iya,Brown:2023mqi,Chicherin:2024hes}.

The remainder of this paper is organised as follows. In Section \ref{sec:kinematics} we review the kinematics of scattering amplitudes in the split-signature dual-momentum space. In Section \ref{sec:amp_review} we review the definition of the amplituhedron for MHV integrands in dual momentum space. In Section \ref{sec:graphical} we introduce the graphical notation which will be used throughout the paper. To gain familiarity with this new notation we provide several pre-existing formulae for the one and two-loop integrands in graphical notation, including the chamber, Kermit and chiral box expansions of the one-loop integrand, together with the fibration formula for the two-loop integrand. In Section \ref{sec:ladders} study the negative ladders in loop space. It is in this section we present the main result of this paper: the formula \eqref{eq:main_result} for all ladders in loop space for any number of particles. In Section \ref{sec:conc} we conclude with an outlook to future research directions.
\section{Kinematics}
\label{sec:kinematics}
We will be working in the dual-momentum space $\mathbb{R}^{2,2}$ with signature $(+,+,-,-)$. The scattering data for $n$-point massless amplitudes is given by a set of $n$ four-dimensional momenta $p_i^\mu$, $i=1,\ldots,n$, $\mu=1,2,3,4$, subject to momentum conservation 
\begin{align}
\sum_{i=1}^n p^\mu_i=0,
\label{eq:dual_coords}
\end{align}
and the massless on-shell condition $p^2=0$. The momentum conservation condition can be trivialised by introducing dual momentum coordinates $x_i^\mu$ defined as 
\begin{align}
p_i^\mu:=x_{i+1}^\mu-x_i^\mu.
\end{align}
Through this relation the inflowing momenta of the scattering process specify the edges of a polygon in $\mathbb{R}^{2,2}$ with vertices ${\bf x} := (x_1,\ldots,x_{n})$ whose consecutive vertices $x_i$ and $x_{i+1}$ are null separated. We denote the set of all null polygons in $\mathbb{R}^{2,2}$ with $n$ vertices as $\mathcal{P}_n$ such that ${\bf x} \in \mathcal{P}_n$. The definition of the dual momentum coordinates is invariant under a global translation and it is convenient to make the choice $x_1=0$ allowing us to invert relation \eqref{eq:dual_coords} to find
\begin{align}
x_j^\mu = \sum_{i=1}^{j-1} p_i^\mu.
\label{eq:invert_dual}
\end{align}
Given a point $x\in\mathbb{R}^{2,2}$ we can split $\mathbb{R}^{2,2}$ into points null separated $(y-x)^2=0$, positively separated $(y-x)^2>0$ and negatively separated $(y-x)^2<0$ from $x$ where the distance between two points $x$ and $y$ is given by
\begin{align}
(x-y)^2=(x^1-y^1)^2+(x^2-y^2)^2-(x^3-y^3)^2-(x^4-y^4)^2.
\end{align}
A concept which will play an important role when coming to study the structure of the amplituhedron is the notion of {\it quadruple cut points}. Given four points $x_i$, $x_j$, $x_k$ and $x_l$ generically there exist two points $q^\pm_{ijkl}$, which we refer to as quadruple cut points, satisfying the quadruple cut conditions
\begin{align}
(q_{ijkl}^\pm-x_i)^2=(q_{ijkl}^\pm-x_j)^2=(q_{ijkl}^\pm-x_k)^2=(q_{ijkl}^\pm-x_l)^2=0,
\end{align} 
where the two solutions are distinguished by the following equation
\begin{align}
\text{sgn} \left| \begin{matrix}
1 & 1 & 1 & 1 & 1 \\ 
x_i & x_j & x_k & x_l & q^\pm_{ijkl}
\end{matrix} \right| = \pm 1.
\end{align}
As our focus will be on MHV integrands it is useful to introduce the notation
\begin{equation}\label{eq:lij}
\begin{cases}
\ell_{ij}^*=q_{ii+1jj+1}^+, \ ,&|i-j|>1,\\
 \tilde{\ell}_{ij}^*=q_{ii+1jj+1}^- \ ,&|i-j|>1,
 \\ \ell^*_{ii+1}=\tilde{\ell}_{ii+1}^*=x_{i+1}\,.
\end{cases}
\end{equation}
The massless on-shell condition, $p^2=0$, can be resolved via the introduction of spinor-helicity variables as
\begin{align}
p^{\alpha \dot{\alpha}}=\left(\begin{matrix} p^0+p^2 & p^1+p^3\\-p^1+p^3&p^0-p^2 \end{matrix}\right) = \lambda^\alpha \tilde{\lambda}^{\dot\alpha},
\label{eq:spinor_helicity}
\end{align}
where $\alpha=1,2$, $\dot{\alpha}=1,2$, and $\lambda,\tilde{\lambda}$ are real variables defined up to the little group rescaling $\lambda \rightarrow t \lambda$,  $\tilde \lambda \rightarrow t^{-1} \tilde\lambda$ for $t \in \mathbb{R}$. Therefore, in the spinor-helicity formalism, each scattering process is determined by a pair of $2 \times n$ matrices $(\lambda,\tilde{\lambda})$ constrained to be orthogonal to one another $\lambda \tilde{\lambda}^T=0$ due to momentum conservation. We denote the set of all such pairs $(\lambda,\tilde{\lambda})$ as $\mathcal{K}_n$. In the spinor-helicity formalism the following brackets appear frequently
\begin{align}
\langle ij \rangle:= \lambda^1_i \lambda^2_j - \lambda^2_i \lambda^1_j, \quad \quad  [ij] := \tilde{\lambda}^{\dot1}_i \tilde{\lambda}^{\dot2}_j - \tilde{\lambda}^{\dot2}_i \tilde{\lambda}^{\dot1}_j.  
\end{align}
\section{Amplituhedron Review}
\label{sec:amp_review}
Throughout this paper we will use the definition of the loop momentum amplituhedron $\mathcal{M}_{n,k,L}$ directly in the split signature dual momentum space $\mathbb{R}^{2,2}$, as defined in \cite{Ferro:2023qdp}. At tree-level the momentum amplituhedron $\mathcal{M}_{n,k,0}$, whose precise definition we do not need but can be found in \cite{Damgaard:2019ztj}, is defined as a subset of the spinor-helicity kinematic space $\mathcal{M}_{n,k,0}\subset \mathcal{K}_n$. Using relations \eqref{eq:invert_dual} and \eqref{eq:spinor_helicity} points $(\lambda,\tilde{\lambda}) \in \mathcal{K}_n$ can be mapped to the space of null polygons $\mathcal{P}_n$ in $\mathbb{R}^{2,2}$ as
\begin{align}
\mathcal{K}_n \ni (\lambda,\tilde\lambda) \mapsto {\bf x}_{(\lambda,\tilde{\lambda})} \in \mathcal{P}_n.
\end{align}
We will be interested in the subset of null polygons obtained as images of points $(\lambda,\tilde{\lambda}) \in \mathcal{M}_{n,k,0}$ which are relevant to the N$^{(k-2)}$MHV$_n$ amplitude. We denote the set of null polygons obtained as images of $(\lambda,\tilde{\lambda}) \in \mathcal{M}_{n,k,0}$ as $\mathcal{P}_{n,k}$. As described in \cite{Ferro:2023qdp} the set of null polygons $\mathcal{P}_{n,k}$ can be characterised without reference to the definition of the momentum amplituhedron as follows: for fixed $(\lambda,\tilde\lambda)\in \mathcal{K}_n$ such that
\begin{itemize}
\item all consecutive brackets of $\lambda$ are positive $\langle ii+1\rangle>0$,
\item and the sequences of brackets
$$
\{\langle i\,i+1\rangle,\langle i\,i+2\rangle,\ldots,\langle i\,i-1\rangle\},
$$
have $k-2$ sign flips for all $i=1,\ldots,n$,
\end{itemize}
a null polygon $\mathbf{x}_{(\lambda,\tilde\lambda)}$ is inside $\mathcal{P}_{n,k}$ if its vertices satisfy the following conditions:
\begin{itemize}
\item all non-consecutive vertices of $\mathbf{x}_{(\lambda,\tilde\lambda)}$ are positively separated
\begin{equation}
(x_i-x_j)^2> 0 \text{ for all } |i-j|>1,
\end{equation}
\item and the sequences of distances
$$\{\langle i+1\,i+2\rangle(x_i-\ell^*_{i+1\,i+2})^2,\langle i+1\,i+3\rangle(x_i-\ell^*_{i+1\,i+3})^2,\ldots,\langle i+1\,i-2\rangle(x_i-\ell^*_{i+1\,i-2})^2\},$$
 have $k-2$ sign flips for all $i=1,\ldots,n$.
 We pick up a factor of $(-1)^{k-1}$ for $\langle ij\rangle $ when $j>n$ due to the twisted cyclic symmetry.
\end{itemize}
At loop level, points inside the momentum amplituhedron $\mathcal{M}_{n,k,L}$ \cite{Ferro:2022abq} are specified by a point $(\lambda,\tilde{\lambda}) \in \mathcal{M}_{n,k,0}$ together with a collection of $L$ loop momenta $\ell_a$. After translating into dual momentum space, the loop momentum amplituhedron is specified by points $(\lambda,{\bf x})$, where ${\bf x}$ is a null polygon ${\bf x} \in \mathcal{P}_{n,k}$, together with a collection of loop dual momenta $y_a$, for $a=1,\ldots,L$, satisfying additional positivity constraints:
\begin{itemize}
	\item for each $y_a$, the distances between $y_a$ and all vertices $x_i$ of the null polygon are non-negative
	\begin{equation}\label{eq:dist-non-neg}
		(y_a-x_i)^2\geq 0, \qquad \text{for all } i=1,\ldots,n\,,
	\end{equation}
	\item for each $y_a$, the sequences
	\begin{equation}\label{eq:signflipsforloops}
		\{\langle i\,i+1\rangle(y_a-\ell^*_{i\,i+1})^2,\langle i\,i+2\rangle(y_a-\ell^*_{i\,i+2})^2,\ldots,\langle i\,i+n-1\rangle(y_a-\ell^*_{i\,i+n-1})^2\}\,,
	\end{equation} 
	have $k$ sign flips for all $i=1,\ldots,n$. 
	\item for each pair of loop momenta $y_a$ and $y_b$ the distance between them is non-negative
	\begin{equation}
	(y_a-y_b)^2\geq 0, \qquad \text{for all } a,b=1,\ldots,L\,.
	\end{equation}
\end{itemize}

\subsection{Chambers}
In this paper, we focus solely on MHV amplitudes which fixes $k=2$ in the remaining sections. The definition of the momentum amplituhedron simplifies in this case. In particular, to define points in the momentum amplituhedron $\mathcal{M}_{n,2,L}$, the matrix $\lambda$ is a $2\times n$ positive matrix, i.e. all its $2\times 2$ minors are positive which implies all distances $(x_i-\ell_{jk}^*)^2$ are positive. Viewing the tree-level data as fixed the loop momentum amplituhedron constrains each loop variable $y_a$, $a=1,2,\ldots,L$ to a compact region $\Delta(\mathbf x)\subset\mathbb{R}^{2,2}$ which we refer to as the {\it one-loop fiber}. The combinatorial structure of the one-loop fibers were studied in \cite{Ferro:2023qdp} for arbitrary helicity, however, we will be interested only in the $k=2$ case where it was found, independent of the tree-level data chosen, that the vertices of $\Delta(\mathbf x)$ are given by the vertices $x_i$ of the null polygon together with the quadruple cut points $\ell_{ij}^*$ for $1<i<j<n$. Also, in \cite{Ferro:2023qdp} explicit expressions for the canonical forms of the one-loop fibers were provided resulting in a novel re-writing of the one-loop integrand which was referred to as a {\it fibration over tree-level} as
\begin{align}
\Omega_{{\bf x};y_1} \left[ \mathcal{M}_{n,2,1} \right]= \Omega_{{\bf x}} \left[ \mathcal{M}_{n,2,0} \right] \wedge \Omega_{y_1}[\Delta({\bf x})]\,,
\end{align}
where $\Omega_{{\bf x};y_1} \left[ \mathcal{M}_{n,2,1} \right]$ denotes the canonical form of the one-loop momentum amplituhedron, $\Omega_{{\bf x}} \left[ \mathcal{M}_{n,2,0} \right]$ is the canonical form of the tree-level momentum amplituhedron and $\Omega_{y_1}[\Delta({\bf x})]$  denotes the canonical form for the one-loop fiber. We leave detailed expressions for the canonical form of the one-loop fiber until the next section. 

Recently, this fibration formula was extended to the two-loop integrand in \cite{Ferro:2024vwn}. At two loops, again viewing the tree-level data as fixed, the loop momentum amplituhedron consists of two points $y_1$ and $y_2$ both constrained to the one-loop fiber $\Delta({\bf x})$, and positively separated from each other. If we further fix $y_1 \in \Delta({{\bf x}})$ then the point $y_2$ is constrained to a subregion $\Delta^+({\bf x};y_1)$ of the one-loop fiber $\Delta({\bf x})$ which is positively separated from $y_1$. We refer to the subregion $\Delta^+({\bf x};y_1)$ to which $y_2$ is constrained as the {\it positive two-loop fiber}. It is important to note that two points $y_1,y_1' \in \Delta({\bf x})$ can result in positive two-loop fibers $\Delta^+({\bf x};y_1)$ and $\Delta^+({\bf x};y_1')$ whose combinatorial structures differ. This is in contrast to the case of tree-level MHV amplitudes where, independent of the tree-level data chosen, the combinatorial structure of the one-loop fiber remains unchanged. This observation motivated \cite{Ferro:2024vwn} to introduce the notion of {\it one-loop chambers} $\mathfrak{c}_{\alpha}$, defined as subregions of the one-loop fiber for which any two-points $y_1,y_1' \in \mathfrak{c}_{\alpha}$ lead to positive two-loop fibers with the {\it same} combinatorial structure. We introduce the notation $\Delta^+_{\alpha}({\bf x};y_1)$ for the positive two-loop fiber associated to the one-loop chamber $\mathfrak{c}_{\alpha}$. As was also pointed out in \cite{Ferro:2024vwn}, there exists an alternative definition of one-loop chambers. Given a point $y_1 \in \Delta({\bf x})$ we can compute the set of signs for the distances $\text{sgn} (y_1-\ell^*_{ij})^2$, for $1<i<j<n$. Then two points $y_1,y_1' \in \Delta({\bf x})$ are in the same one-loop chamber if they produce the same sign pattern. Therefore, the problem of determining the set of one-loop chambers reduces to the problem of determining the set of all possible sign patterns for the $\text{sgn} (y_1-\ell^*_{ij})^2$. This decomposition is closely related to the decomposition of the $m=2$ amplituhedron and the hypersimplex, as explained in \cite{Parisi:2021oql}. Importantly, the set of one-loop chambers are known to be counted by the Eulerian numbers $n_{\mathfrak{C}}=E_{3,n-1}$. Using the decomposition of the one-loop fiber into one-loop chambers, the two-loop momentum amplituhedron can be written in the following {\it fibration of fibration} form
\begin{align}
\Omega_{{\bf x};y_1y_2} \left[ \mathcal{M}_{n,2,1} \right]= \Omega_{{\bf x}} \left[ \mathcal{M}_{n,2,0} \right] \wedge \sum_{\alpha \in [n_{\mathfrak{C}}]} \Omega_{y_1}[\mathfrak{c}_{\alpha}] \wedge \Omega_{y_2}[\Delta^+_{\alpha}({\bf x};y_1)].
\end{align}
As we will see in the next section a detailed knowledge of the chamber structure is not needed in order to write down any of the forms we present in this paper, therefore, we do not present explicit expressions for their forms. 

An interesting question to ask is how this fibration of fibration picture can be extended to higher loop geometries. Whereas we expect this to be a difficult task in general, there exist a simpler set of geometries, which we refer to as ladders in loop space, defined by relaxing or flipping positivity constraints between the loop momenta, which provide a perfect application for the fibration of fibration picture. We will define the ladder geometries in detail in the next section, for now we introduce the three-loop positive ladder which is defined by relaxing the positivity constraint $(y_1-y_3)^2$ in the definition of the momentum amplituhedron. Recall that in order to find explicit canonical forms for two-loop momentum amplituhedron we decompose the one-loop fiber into one-loop chambers and multiply each chamber by its corresponding positive two-loop fiber. Now, in order to extend this idea to three-loop positive ladder, we decompose each positive fiber $\Delta_{\alpha}^+({\bf x};y_1)$ into one-loop chambers $\mathfrak{c}_{\beta}$ and multiply each chamber by its corresponding positive fiber $\Delta_{\beta}^+({\bf x};y_2)=\{y_3\in\Delta({\mathbf x}):(y_3-y_2)^2>0\}$ which reads\footnote{For the remainder of the paper we have stripped of the tree-level factor $\Omega_{{\bf x}}[\mathcal{M}_{n,2,0}]$ from all forms.}
\begin{align}
\begin{tikzpicture}[baseline=-0.5ex,scale=0.7]
        \draw[ultra thick,green] (0,0) -- (2,0);
        \node[fill=black, circle, minimum size=4pt, inner sep=0pt, line width=1pt, draw=black] (A) at (0,0) {};
        \node[fill=black, circle, minimum size=4pt, inner sep=0pt, line width=1pt, draw=black] (B) at (1,0) {};
        \node[fill=black, circle, minimum size=4pt, inner sep=0pt, line width=1pt, draw=black] (C) at (2,0) {};
         \node[below=4pt] at (0,0) {$y_1$};
         \node[below=4pt] at (1,0) {$y_2$};
         \node[below=4pt] at (2,0) {$y_{3}$};
\end{tikzpicture}  =  \sum_{\alpha \in [n_{\mathfrak{C}}]} \Omega_{y_1}[\mathfrak{c}_{\alpha}] &\wedge \sum_{\beta \in [n_{\mathfrak{C}}]} \Omega_{y_2}[\Delta^+_{\alpha}({\bf x};y_1) \cap \mathfrak{c}_{\beta}]    \wedge \Omega_{y_{3}}[\Delta^+_{\beta}({\bf x};y_{2}) ].
\end{align}
We now move on to study this procedure in detail for ladder geometries at all loops. We begin by introducing a graphical notation for the ladder geometries and one-loop chambers. This will include formulae for the one and two-loop momentum amplituhedron forms, as well as the positive and negative two-loop fibers in terms of chiral boxes. As was already emphasised, we will not need the full knowledge of the chamber structure in order to write any of the canonical forms.
\section{Graphical Notation}
\label{sec:graphical}
The positivity constraints between loop momenta play an important role in the definition of the amplituhedron. As such, it is useful to introduce a notation that keeps track of them. With this in mind we follow the notation originally introduced in the {\it negative geometries} literature \cite{Arkani-Hamed:2021iya}. For each loop momentum $y_a$ we introduce a circular vertex which represents the conditions constraining the loop momentum to the one-loop fiber. In other words the canonical form of the one-loop fiber is represented as
\begin{align}
\begin{tikzpicture}[baseline=-0.5ex,scale=0.8]
        \draw[ultra thick,green] (0,0);
         \node[below=4pt] at (0,0) {$y_a$};
        \node[fill=black, circle, minimum size=5pt, inner sep=0pt, line width=1pt, draw=black] (A) at (0,0) {};
\end{tikzpicture} =\Omega_{y_a} \left[ \Delta({\bf x}) \right].
\end{align}
To encode positivity $(y_a-y_b)^2>0$ or negativity $(y_a-y_b)^2<0$ between loop momenta we introduce a green or red edge between vertices $y_a$ and $y_b$ respectively such that the $L$-loop amplithuedron form is represented by the complete graph on $L$ vertices with all green edges. We follow the terminology introduced in \cite{Brown:2023mqi} and refer to graphs constructed in this way as {\it graphs in loop space}. As an example, the two-loop momentum amplituhedron form corresponds to the two-loop positive ladder
\begin{align}
\begin{tikzpicture}[baseline=-0.5ex,scale=0.8]
        \draw[ultra thick,green] (0,0) -- (1,0);
         \node[below=4pt] at (0,0) {$y_1$};
         \node[below=4pt] at (1,0) {$y_2$};
        \node[fill=black, circle, minimum size=5pt, inner sep=0pt, line width=1pt, draw=black] (A) at (0,0) {};
        \node[fill=black, circle, minimum size=5pt, inner sep=0pt, line width=1pt, draw=black] (B) at (1,0) {};
\end{tikzpicture} = \Omega_{y_1y_2}[\{ y_1,y_2 \in \Delta({\bf x})  :  (y_1-y_2)^2> 0 \}].
\end{align}
We now wish to extend this notation further as to include the notion of a one-loop chamber. Recall, the one-loop chambers are defined by fixing a sign pattern for the distances $(y_a-\ell^*_{ij})^2$. Therefore, in order to encapsulate these additional constraints, we introduce {\it square} vertices for each of the $\ell^*_{ij}$, where positivity/negativity of the distance $(y_a-\ell^*_{ij})^2$ is again encoded by a green/red edge between the vertices $y_a$ and $\ell^*_{ij}$ such that we have for example\footnote{Since the loop momenta automatically satisfy $(y-x_i)^2>0$ we omit the vertices labelled by the $x_i$.}
\begin{align}
\begin{tikzpicture}[baseline=-0.5ex,scale=0.8]
        \draw[ultra thick,green]  (-1,0)--(0,0);
        \node[below=4pt] at (-1,0) {$y_a$};
        \node[below=4pt] at (0,0) {$\ell^*_{ij}$};
         \node[rectangle, draw, fill=black, minimum width=5 pt, minimum height=5 pt,inner sep=0pt] (pt2) at (0,0) {};
        \node[fill=black, circle, minimum size=5pt, inner sep=0pt, line width=1pt, draw=black] at (-1,0) {};
\end{tikzpicture} &= \Omega_{y_a}[\{ y_a \in \Delta({\bf x}) : (y_a-\ell^*_{ij})^2>0 \}], \notag \\
 \begin{tikzpicture}[baseline=-0.5ex,scale=0.8]
        \draw[ultra thick,red]  (-1,0)--(0,0);
        \node[below=4pt] at (-1,0) {$y_a$};
        \node[below=4pt] at (0,0) {$\ell^*_{ij}$};
         \node[rectangle, draw, fill=black, minimum width=5 pt, minimum height=5 pt,inner sep=0pt] (pt2) at (0,0) {};
        \node[fill=black, circle, minimum size=5pt, inner sep=0pt, line width=1pt, draw=black] at (-1,0) {};
\end{tikzpicture} &= \Omega_{y_a}[\{ y_a \in \Delta({\bf x}) : (y_a-\ell^*_{ij})^2<0 \}].
\label{eq:graph_regions}
\end{align} 
We note that square vertices are used for the $\ell^*_{ij}$ to indicate that these are viewed as fixed points and as such the above is {\it not} a canonical form in these variables. In the remainder of the paper we will refer to square vertices as being {\it frozen}. Since the one-loop chambers $\mathfrak{c}_\alpha$ are defined by fully fixing the sign pattern for all distances $(y-\ell^*_{ij})^2$ it is useful to introduce the sets
\begin{align}
\vertsGreen{\alpha} &= \{ \ell^*_{ij} \ | \text{ for all }  y \in \mathfrak{c}_{\alpha} \text{ we have } (y-\ell^*_{ij})^2>0 \} \cup  \{ x_1,\ldots,x_n\}, \notag \\
\vertsRed{\alpha} &= \{ \ell^*_{ij}  | \text{ for all }  y \in \mathfrak{c}_{\alpha} \text{ we have }  (y-\ell^*_{ij})^2<0 \},
\end{align}
and for $n>4$ the sets\footnote{For $n=4$ we have $\mathcal{V}^+ = \{ x_1,\ldots,x_4\}$ and  $\mathcal{V}^- = \{ \ell_{13}^*,\ell_{24}^*\}$.}
\begin{align}
\mathcal{V}^+ = \{ \ell^*_{ij }\} \cup  \{ x_1,\ldots,x_n\}, \quad \quad \mathcal{V}^- = \{ \ell^*_{ij }\}.
\end{align}
In order to simplify our graphical notation, we allow frozen vertices to be labelled by sets of points. If a vertex is labelled by a set of points $\mathcal{U}\subset\mathcal{V}_{\alpha}^{\pm}$, then any green/red link attached to this vertex indicates that all vertices $\ell^*_{ij} \in \mathcal{U}$ are positively/negatively separated from the point labelling the other end of the link. With this notation the canonical form for the one-loop chambers are depicted as 
\begin{align}
  \Omega_{y_a}[\mathfrak{c}_\alpha]= \begin{tikzpicture}[baseline=-0.5ex,scale=0.8]
        \draw[ultra thick,green]  (-1,0)--(0,0.5);
        \draw[ultra thick,red]  (-1,0)--(0,-0.5);
        \node[below=4pt] at (-1,0) {$y_a$};
        \node[right=4pt] at (0,0.5) {$\vertsGreen{\alpha}$};
        \node[right=4pt] at (0,-0.5) {$\vertsRed{\alpha}$};
        \node[rectangle, draw, fill=black, minimum width=5 pt, minimum height=5 pt,inner sep=0pt] (pt2) at (0,0.5) {};
        \node[rectangle, draw, fill=black, minimum width=5 pt, minimum height=5 pt,inner sep=0pt] (pt2) at (0,-0.5) {};
        \node[fill=black, circle, minimum size=5pt, inner sep=0pt, line width=1pt, draw=black] at (-1,0) {};
\end{tikzpicture}.
\label{eq:graphical_chamber}
\end{align}
It is important to note that in the next section we will see instances of graphs where one of the loop momenta become frozen, for example the canonical forms for the positive two-loop fibers of the last section are represented as 
\begin{align}
\Omega_{y_2}[\Delta_{\alpha}^+({\bf x};y_1)]= \Omega_{y_2}[\{ y_1 \in \mathfrak{c}_\alpha,y_2 \in \Delta({\bf x}) : (y_1-y_2)^2> 0 \}]=\begin{tikzpicture}[baseline=-0.5ex,scale=0.8]
        \draw[ultra thick,green] (0,0) -- (1,0);
        \draw[ultra thick,green]  (-1,0.5)--(0,0);
        \draw[ultra thick,red]  (-1,-0.5)--(0,0);
        \node[left=4pt] at (-1,0.5) {$\vertsGreen{\alpha}$};
        \node[left=4pt] at (-1,-0.5) {$\vertsRed{\alpha}$};
         \node[below=4pt] at (0,0) {$y_1$};
         \node[below=4pt] at (1,0) {$y_2$};
         \node[rectangle, draw, fill=black, minimum width=5 pt, minimum height=5 pt,inner sep=0pt] (pt1) at (-1,0.5) {};
         \node[rectangle, draw, fill=black, minimum width=5 pt, minimum height=5 pt,inner sep=0pt] (pt2) at (-1,-0.5) {};
        \node[fill=black, circle, minimum size=5pt, inner sep=0pt, line width=1pt, draw=black] at (1,0) {};
        \node[rectangle, draw, fill=black, minimum width=5 pt, minimum height=5 pt,inner sep=0pt] (pt1) at (0,0) {};
\end{tikzpicture} .
\end{align}
Here $y_1$ is treated as a fixed point inside the one-loop chamber $\mathfrak{c}_{\alpha}$ and hence the above is a canonical form only in $y_2$. As a final piece of notation we introduce the following subset of chambers
\begin{align}
\mathfrak{C}^\pm_{ij} = \{ \alpha \in[  n_{\mathfrak{C}}] : \ell^*_{ij} \in \mathcal{V}^\pm_{\alpha} \},
\end{align}
namely $\mathfrak{C}^+_{ij}$ (resp. $\mathfrak{C}^-_{ij}$) contains all chambers for which $\ell_{ij}^*$ is positively (resp. negatively) separated from all points in a chamber.
A frequently used trick that will be used throughout the paper will be to perform the following reorganisation of sums
\begin{align}
\sum_{\alpha \in [n_{\mathfrak{C}}]}\sum_{\ell^*_{ij} \in \mathcal{V}^\pm_{\alpha}} = \sum_{\ell^*_{ij} \in \mathcal{V}^\pm } \sum_{\alpha \in \mathfrak{C}_{ij}^\pm }.
\label{eq:sum_trick}
\end{align}
\subsection{One-loop Geometries}
To gain familiarity with this new notation it is useful to see various known formulae for the one-loop integrand in this notation. We begin with the chamber decomposition presented in \cite{Ferro:2024vwn} which reads
\begin{align}
\begin{tikzpicture}[baseline=-0.5ex,scale=0.8]
         \node[below=4pt] at (0,0) {$y$};
        \node[fill=black, circle, minimum size=5pt, inner sep=0pt, line width=1pt, draw=black] (A) at (0,0) {};
\end{tikzpicture} & =\sum_{\alpha \in [n_{\mathfrak{C}}]} \begin{tikzpicture}[baseline=-0.5ex,scale=0.8]
        \draw[ultra thick,green]  (-1,0.5)--(-2,0);
        \draw[ultra thick,red]  (-1,-0.5)--(-2,0);
        \node[right=4pt] at (-1,0.5) {$\vertsGreen{\alpha}$};
        \node[right=4pt] at (-1,-0.5) {$\vertsRed{\alpha}$};
         \node[below=4pt] at (-2,0) {$y$};
         \node[rectangle, draw, fill=black, minimum width=5 pt, minimum height=5 pt,inner sep=0pt] (pt1) at (-1,0.5) {};
         \node[rectangle, draw, fill=black, minimum width=5 pt, minimum height=5 pt,inner sep=0pt] (pt2) at (-1,-0.5) {};
        \node[fill=black, circle, minimum size=5pt, inner sep=0pt, line width=1pt, draw=black] at (-2,0) {};
\end{tikzpicture}.
\label{eq:chamb_decomp}
\end{align}
The terms appearing on the right hand side of \eqref{eq:chamb_decomp}  are the one-loop chamber forms. Importantly, we will not need their explicit expressions in this paper. Next, we consider the Kermit expansion of \cite{Arkani-Hamed:2010zjl} which reads 
\begin{align}
\begin{tikzpicture}[baseline=-0.5ex,scale=0.8]
         \node[below=4pt] at (0,0) {$y$};
        \node[fill=black, circle, minimum size=5pt, inner sep=0pt, line width=1pt, draw=black] (A) at (0,0) {};
\end{tikzpicture}  =\sum_{1<i<j<n} \begin{tikzpicture}[baseline=-0.5ex,scale=0.8]
        \draw[ultra thick,green]  (-1,0.5)--(-2,0);
        \draw[ultra thick,red]  (-1,-0.5)--(-2,0);
        \node[right=4pt] at (-1,0.5) {$\mathcal{K}_{1ij}^+$};
        \node[right=4pt] at (-1,-0.5) {$\mathcal{K}_{1ij}^-$};
         \node[below=4pt] at (-2,0) {$y$};
         \node[rectangle, draw, fill=black, minimum width=5 pt, minimum height=5 pt,inner sep=0pt] (pt1) at (-1,0.5) {};
         \node[rectangle, draw, fill=black, minimum width=5 pt, minimum height=5 pt,inner sep=0pt] (pt2) at (-1,-0.5) {};
        \node[fill=black, circle, minimum size=5pt, inner sep=0pt, line width=1pt, draw=black] at (-2,0) {};
\end{tikzpicture}.
\end{align}
where the sum runs over the so-called Kermit cells, and the subsets $\mathcal{K}^\pm_{1ij}$ are defined as
\begin{align}
\mathcal{K}^+_{1ij} &= \{\ell_{13}^* \ldots \ell_{1i}^* \}\cup \{\ell_{1j+1}^* \ldots \ell_{1n}^* \} \cup \{ x_1,\ldots,x_n\}, \notag \\
\mathcal{K}^-_{1ij} &= \{\ell_{1i+1}^*\ldots \ell_{1j}^*\}.
\end{align}
Expressions for the canonical forms of the Kermit cells can be found in \cite{Arkani-Hamed:2010zjl}. Finally, the last expansion we wish to introduce a graphical notation for is the chiral box expansion of \cite{Arkani-Hamed:2010pyv}. For this we introduce the notation
\begin{align}
\begin{tikzpicture}[baseline=-0.5ex,scale=0.8]
         \node[below=4pt] at (0,0) {$y$};
        \node[fill=black, circle, minimum size=5pt, inner sep=0pt, line width=1pt, draw=black] (A) at (0,0) {};
\end{tikzpicture} =
  \sum_{x_i}  \begin{tikzpicture}[baseline=-0.5ex,scale=0.8]
        \draw[ultra thick,black] (0,0) -- (1,0);
        \draw[ultra thick,black]  (-1,0)--(0,0);
        \node[below=4pt] at (-1,0) {$x_i$};
         \node[below=4pt] at (0,0) {$*$};
         \node[below=4pt] at (1,0) {$y$};
          \node[rectangle, draw, fill=black, minimum width=5 pt, minimum height=5 pt,inner sep=0pt] (pt1) at (-1,0) {};
        \node[rectangle, draw, fill=black, minimum width=5 pt, minimum height=5 pt,inner sep=0pt] (A) at (0,0) {};
        \node[fill=black, circle, minimum size=5pt, inner sep=0pt, line width=1pt, draw=black] (B) at (1,0) {};
\end{tikzpicture}+  \sum_{\ell^*_{ij}}  \begin{tikzpicture}[baseline=-0.5ex,scale=0.8]
        \draw[ultra thick,black] (0,0) -- (1,0);
        \draw[ultra thick,black]  (-1,0)--(0,0);
        \node[below=4pt] at (-1,0) {$\ell^*_{ij}$};
         \node[below=4pt] at (0,0) {$*$};
         \node[below=4pt] at (1,0) {$y$};
          \node[rectangle, draw, fill=black, minimum width=5 pt, minimum height=5 pt,inner sep=0pt] (pt1) at (-1,0) {};
        \node[rectangle, draw, fill=black, minimum width=5 pt, minimum height=5 pt,inner sep=0pt] (A) at (0,0) {};
        \node[fill=black, circle, minimum size=5pt, inner sep=0pt, line width=1pt, draw=black] (B) at (1,0) {};
\end{tikzpicture},
\end{align}
where $*$ is an arbitrary point in dual-momentum space and we have defined the chiral box integrands as 
\begin{align}
\begin{tikzpicture}[baseline=-0.5ex,scale=0.8]
        \draw[ultra thick,black] (0,0) -- (1,0);
        \draw[ultra thick,black]  (-1,0)--(0,0);
        \node[below=4pt] at (-1,0) {$x_i$};
         \node[below=4pt] at (0,0) {$*$};
         \node[below=4pt] at (1,0) {$y$};
          \node[rectangle, draw, fill=black, minimum width=5 pt, minimum height=5 pt,inner sep=0pt] (pt1) at (-1,0) {};
        \node[rectangle, draw, fill=black, minimum width=5 pt, minimum height=5 pt,inner sep=0pt] (A) at (0,0) {};
        \node[fill=black, circle, minimum size=5pt, inner sep=0pt, line width=1pt, draw=black] (B) at (1,0) {};
\end{tikzpicture} & := \frac{4(x_{i-1}-x_{i+1})^2(x_i-*)^2}{(y-x_{i-1})^2(y-x_{i})^2(y-x_{i+1})^2(y-*)^2} \notag, \\
\begin{tikzpicture}[baseline=-0.5ex,scale=0.8]
        \draw[ultra thick,black] (0,0) -- (1,0);
        \draw[ultra thick,black]  (-1,0)--(0,0);
        \node[below=4pt] at (-1,0) {$\ell^*_{ij}$};
         \node[below=4pt] at (0,0) {$*$};
         \node[below=4pt] at (1,0) {$y$};
          \node[rectangle, draw, fill=black, minimum width=5 pt, minimum height=5 pt,inner sep=0pt] (pt1) at (-1,0) {};
        \node[rectangle, draw, fill=black, minimum width=5 pt, minimum height=5 pt,inner sep=0pt] (A) at (0,0) {};
        \node[fill=black, circle, minimum size=5pt, inner sep=0pt, line width=1pt, draw=black] (B) at (1,0) {};
\end{tikzpicture} &:= \frac{4S_{x_ix_{i+1}x_jx_{j+1}}(y-\tilde{\ell}^*_{ij})^2(\ell^*_{ij}-*)^2}{(y-x_i)^2(y-x_{i+1})^2(y-x_j)^2(y-x_{j+1})^2(y-*)^2},
\label{eq:chiral_box}
\end{align}
where
\begin{align}
S_{x_ix_{i+1}x_jx_{j+1}}= (x_i-x_{j+1})^2+(x_{i+1}-x_j)^2-(x_i-x_j)^2-(x_{i+1}-x_{j+1})^2.
\end{align}
Importantly, we do not color the links in graphs in \eqref{eq:chiral_box} since they rather provide a label for a chiral pentagon differential form and do not correspond to any geometry defined by positive/negative distances.

\subsubsection{Four Points}
To get a better understanding of the formulae above, it is useful to look at a couple of low point examples, starting with four points. At four-point the one-loop region is covered by a single chamber leading to
\begin{align}
\mathcal{V}^+ = \mathcal{V}^+_1 &= \{ x_1,x_2,x_3,x_4\}, && \mathcal{V}^- = \mathcal{V}^-_1 = \{ \ell^*_{13}, \ell^*_{24}\},
\end{align}
where, importantly, the set $\mathcal{V}^+$ does not contain any quadruple cut points. This results in the following graphical identity
\begin{align}
\begin{tikzpicture}[baseline=-0.5ex,scale=0.8]
         \node[below=4pt] at (0,0) {$y_1$};
        \node[fill=black, circle, minimum size=5pt, inner sep=0pt, line width=1pt, draw=black] (A) at (0,0) {};
\end{tikzpicture}=\begin{tikzpicture}[baseline=-0.5ex,scale=0.8]
        \draw[ultra thick,red]  (-1,0)--(-2,0);
        \node[below=4pt] at (-1,0) {{\small $\ell^*_{ij}$}};
        \node[below=4pt] at (-2,0) {{\small $y_1$}};
         \node[rectangle, draw, fill=black, minimum width=5 pt, minimum height=5 pt,inner sep=0pt] (pt1) at (-1,0) {};
        \node[fill=black, circle, minimum size=5pt, inner sep=0pt, line width=1pt, draw=black] at (-2,0) {};
\end{tikzpicture},
\label{eq:simplification_n4}
\end{align}
for $(ij)=(13)$ or $(ij)=(24)$.
It is equivalent to the statement that all points in the one-loop fiber are negatively separated from both quadruple cut points $\ell_{13}^*$ and $\ell_{24}^*$.
\subsubsection{Five Points}
At five-points we begin to see the generic behaviour due to the appearance of chambers. In this case we have $11$ chambers which come in three-cyclic classes defined by 
\begin{align}
\mathcal{V}_1^+ &=\{ x_1,\ldots,x_5 \} \cup \{  \ell^*_{13},\ell^*_{24} \} &&\mathcal{V}_1^-=\{ \ell^*_{35},\ell^*_{14},\ell^*_{25} \}, \notag\\
 \mathcal{V}_6^+ &=\{ x_1,\ldots,x_5 \} \cup \{  \ell^*_{13} \} &&\mathcal{V}_6^-=\{ \ell^*_{24},\ell^*_{35},\ell^*_{14},\ell^*_{25}\}, \notag \\
 \mathcal{V}_{11}^+ &=\{ x_1,\ldots,x_5 \}  &&\mathcal{V}_{11}^-=\{ \ell^*_{13},\ell^*_{24},\ell^*_{35},\ell^*_{14},\ell^*_{25} \}.
\end{align}
Then, the chamber decomposition of the one-loop fiber can be written as 
\begin{align}
\begin{tikzpicture}[baseline=-0.5ex,scale=0.8]
         \node[below=4pt] at (0,0) {$y$};
        \node[fill=black, circle, minimum size=5pt, inner sep=0pt, line width=1pt, draw=black] (A) at (0,0) {};
\end{tikzpicture} & = \left(\begin{tikzpicture}[baseline=-0.5ex,scale=0.8]
        \draw[ultra thick,green]  (-1,0.5)--(-2,0);
        \draw[ultra thick,red]  (-1,-0.5)--(-2,0);
         \draw[ultra thick,green]  (-1,1)--(-2,0);
         \draw[ultra thick,red]  (-1,-1)--(-2,0);
         \draw[ultra thick,red]  (-1,0)--(-2,0);
        \node[right=4pt] at (-1,1) {{\small $\ell^*_{13}$}};
        \node[right=4pt] at (-1,0.5) {{\small $\ell^*_{24}$}};
        \node[right=4pt] at (-1,0) {{\small $\ell^*_{35}$}};
        \node[right=4pt] at (-1,-0.5) {{\small $\ell^*_{14}$}};
        \node[right=4pt] at (-1,-1) {{\small $\ell^*_{25}$}};
        \node[below=4pt] at (-2,0) {{\small $y$}};
         \node[rectangle, draw, fill=black, minimum width=5 pt, minimum height=5 pt,inner sep=0pt] (pt1) at (-1,0.5) {};
         \node[rectangle, draw, fill=black, minimum width=5 pt, minimum height=5 pt,inner sep=0pt] (pt2) at (-1,-0.5) {};
         \node[rectangle, draw, fill=black, minimum width=5 pt, minimum height=5 pt,inner sep=0pt] (pt2) at (-1,1) {};
         \node[rectangle, draw, fill=black, minimum width=5 pt, minimum height=5 pt,inner sep=0pt] (pt2) at (-1,-1) {};
         \node[rectangle, draw, fill=black, minimum width=5 pt, minimum height=5 pt,inner sep=0pt] (pt2) at (-1,0) {};
        \node[fill=black, circle, minimum size=5pt, inner sep=0pt, line width=1pt, draw=black] at (-2,0) {};
\end{tikzpicture} +\begin{tikzpicture}[baseline=-0.5ex,scale=0.8]
        \draw[ultra thick,red]  (-1,0.5)--(-2,0);
        \draw[ultra thick,red]  (-1,-0.5)--(-2,0);
         \draw[ultra thick,green]  (-1,1)--(-2,0);
         \draw[ultra thick,red]  (-1,-1)--(-2,0);
         \draw[ultra thick,red]  (-1,0)--(-2,0);
        \node[right=4pt] at (-1,1) {{\small $\ell^*_{13}$}};
        \node[right=4pt] at (-1,0.5) {{\small $\ell^*_{24}$}};
        \node[right=4pt] at (-1,0) {{\small $\ell^*_{35}$}};
        \node[right=4pt] at (-1,-0.5) {{\small $\ell^*_{14}$}};
        \node[right=4pt] at (-1,-1) {{\small $\ell^*_{25}$}};
        \node[below=4pt] at (-2,0) {{\small $y$}};
         \node[rectangle, draw, fill=black, minimum width=5 pt, minimum height=5 pt,inner sep=0pt] (pt1) at (-1,0.5) {};
         \node[rectangle, draw, fill=black, minimum width=5 pt, minimum height=5 pt,inner sep=0pt] (pt2) at (-1,-0.5) {};
         \node[rectangle, draw, fill=black, minimum width=5 pt, minimum height=5 pt,inner sep=0pt] (pt2) at (-1,1) {};
         \node[rectangle, draw, fill=black, minimum width=5 pt, minimum height=5 pt,inner sep=0pt] (pt2) at (-1,-1) {};
         \node[rectangle, draw, fill=black, minimum width=5 pt, minimum height=5 pt,inner sep=0pt] (pt2) at (-1,0) {};
        \node[fill=black, circle, minimum size=5pt, inner sep=0pt, line width=1pt, draw=black] at (-2,0) {};
\end{tikzpicture} + \text{ cyc}\right)+\begin{tikzpicture}[baseline=-0.5ex,scale=0.8]
        \draw[ultra thick,red]  (-1,0.5)--(-2,0);
        \draw[ultra thick,red]  (-1,-0.5)--(-2,0);
         \draw[ultra thick,red]  (-1,1)--(-2,0);
         \draw[ultra thick,red]  (-1,-1)--(-2,0);
         \draw[ultra thick,red]  (-1,0)--(-2,0);
        \node[right=4pt] at (-1,1) {{\small $\ell^*_{13}$}};
        \node[right=4pt] at (-1,0.5) {{\small $\ell^*_{24}$}};
        \node[right=4pt] at (-1,0) {{\small $\ell^*_{35}$}};
        \node[right=4pt] at (-1,-0.5) {{\small $\ell^*_{14}$}};
        \node[right=4pt] at (-1,-1) {{\small $\ell^*_{25}$}};
        \node[below=4pt] at (-2,0) {{\small $y$}};
         \node[rectangle, draw, fill=black, minimum width=5 pt, minimum height=5 pt,inner sep=0pt] (pt1) at (-1,0.5) {};
         \node[rectangle, draw, fill=black, minimum width=5 pt, minimum height=5 pt,inner sep=0pt] (pt2) at (-1,-0.5) {};
         \node[rectangle, draw, fill=black, minimum width=5 pt, minimum height=5 pt,inner sep=0pt] (pt2) at (-1,1) {};
         \node[rectangle, draw, fill=black, minimum width=5 pt, minimum height=5 pt,inner sep=0pt] (pt2) at (-1,-1) {};
         \node[rectangle, draw, fill=black, minimum width=5 pt, minimum height=5 pt,inner sep=0pt] (pt2) at (-1,0) {};
        \node[fill=black, circle, minimum size=5pt, inner sep=0pt, line width=1pt, draw=black] at (-2,0) {};
\end{tikzpicture}.
\end{align}
Explicit expressions for the chamber forms can be found in \cite{Ferro:2024vwn}. Alternatively the one-loop integrand can be expanded into Kermit cells which for the case of five-points reads
\begin{align}
\begin{tikzpicture}[baseline=-0.5ex,scale=0.9]
         \node[below=4pt] at (0,0) {$y$};
        \node[fill=black, circle, minimum size=5pt, inner sep=0pt, line width=1pt, draw=black] (A) at (0,0) {};
\end{tikzpicture} & = \begin{tikzpicture}[baseline=-0.5ex,scale=0.8]
        \draw[ultra thick,red]  (-1,0.3)--(-2,0);
        \draw[ultra thick,red]  (-1,-0.3)--(-2,0);
        \node[right=4pt] at (-1,0.3) {{\small $\ell^*_{13}$}};
        \node[right=4pt] at (-1,-0.3) {{\small $\ell^*_{14}$}};
        \node[below=4pt] at (-2,0) {{\small $y$}};
         \node[rectangle, draw, fill=black, minimum width=5 pt, minimum height=5 pt,inner sep=0pt] (pt1) at (-1,0.3) {};
         \node[rectangle, draw, fill=black, minimum width=5 pt, minimum height=5 pt,inner sep=0pt] (pt2) at (-1,-0.3) {};
        \node[fill=black, circle, minimum size=5pt, inner sep=0pt, line width=1pt, draw=black] at (-2,0) {};
\end{tikzpicture}+\begin{tikzpicture}[baseline=-0.5ex,scale=0.9]
        \draw[ultra thick,green]  (-1,0.3)--(-2,0);
        \draw[ultra thick,red]  (-1,-0.3)--(-2,0);
        \node[right=4pt] at (-1,0.3) {{\small $\ell^*_{13}$}};
        \node[right=4pt] at (-1,-0.3) {{\small $\ell^*_{14}$}};
        \node[below=4pt] at (-2,0) {{\small $y$}};
         \node[rectangle, draw, fill=black, minimum width=5 pt, minimum height=5 pt,inner sep=0pt] (pt1) at (-1,0.3) {};
         \node[rectangle, draw, fill=black, minimum width=5 pt, minimum height=5 pt,inner sep=0pt] (pt2) at (-1,-0.3) {};
        \node[fill=black, circle, minimum size=5pt, inner sep=0pt, line width=1pt, draw=black] at (-2,0) {};
\end{tikzpicture}+\begin{tikzpicture}[baseline=-0.5ex,scale=0.9]
        \draw[ultra thick,red]  (-1,0.3)--(-2,0);
        \draw[ultra thick,green]  (-1,-0.3)--(-2,0);
        \node[right=4pt] at (-1,0.3) {{\small $\ell^*_{13}$}};
        \node[right=4pt] at (-1,-0.3) {{\small $\ell^*_{14}$}};
        \node[below=4pt] at (-2,0) {{\small $y$}};
         \node[rectangle, draw, fill=black, minimum width=5 pt, minimum height=5 pt,inner sep=0pt] (pt1) at (-1,0.3) {};
         \node[rectangle, draw, fill=black, minimum width=5 pt, minimum height=5 pt,inner sep=0pt] (pt2) at (-1,-0.3) {};
        \node[fill=black, circle, minimum size=5pt, inner sep=0pt, line width=1pt, draw=black] at (-2,0) {};
\end{tikzpicture}.
\end{align}
\subsection{Two-loop Geometries}
At two loops the momentum amplituhedron is defined as the set of points $y_1$ and $y_2$, both constrained to the one-loop fiber $\Delta(\bf x)$, with the additional mutual positivity constraint $(y_1-y_2)^2>0$. In terms of our graphical notation this corresponds to the two-loop positive ladder
\begin{align}
\begin{tikzpicture}[baseline=-0.5ex,scale=0.8]
        \draw[ultra thick,green] (0,0) -- (1,0);
         \node[below=4pt] at (0,0) {$y_1$};
         \node[below=4pt] at (1,0) {$y_2$};
        \node[fill=black, circle, minimum size=5pt, inner sep=0pt, line width=1pt, draw=black] (A) at (0,0) {};
        \node[fill=black, circle, minimum size=5pt, inner sep=0pt, line width=1pt, draw=black] (B) at (1,0) {};
\end{tikzpicture} = \Omega_{y_1y_2}[\{ y_1,y_2 \in \Delta({\bf x}) : (y_1-y_2)^2> 0 \}].
\end{align}
Given that we have a decomposition of the one-loop fiber into chambers $\mathfrak{c}_{\alpha}$ it is natural to consider whether this can also be used to decompose the two-loop positive ladder. To achieve this we can define the regions where $y_1$ is constrained to the one-loop chamber $\mathfrak{c}_\alpha$ and $y_2$ is left unconstrained to range over the entire one-loop fiber $\Delta({\bf x})$. By definition these regions are non-overlapping and cover the entire space of the two-loop positive ladder such that summing over one-loop chambers we find
\begin{align}
\begin{tikzpicture}[baseline=-0.5ex,scale=0.8]
        \draw[ultra thick,green] (0,0) -- (1,0);
         \node[below=4pt] at (0,0) {$y_1$};
         \node[below=4pt] at (1,0) {$y_2$};
        \node[fill=black, circle, minimum size=5pt, inner sep=0pt, line width=1pt, draw=black] (A) at (0,0) {};
        \node[fill=black, circle, minimum size=5pt, inner sep=0pt, line width=1pt, draw=black] (B) at (1,0) {};
\end{tikzpicture} = \sum_{\alpha \in [n_{\mathfrak{C}}]}\begin{tikzpicture}[baseline=-0.5ex,scale=0.8]
        \draw[ultra thick,green] (0,0) -- (1,0);
        \draw[ultra thick,green]  (-1,0.5)--(0,0);
        \draw[ultra thick,red]  (-1,-0.5)--(0,0);
        \node[left=4pt] at (-1,0.5) {$\vertsGreen{\alpha}$};
        \node[left=4pt] at (-1,-0.5) {$\vertsRed{\alpha}$};
         \node[below=4pt] at (0,0) {$y_1$};
         \node[below=4pt] at (1,0) {$y_2$};
         \node[rectangle, draw, fill=black, minimum width=5 pt, minimum height=5 pt,inner sep=0pt] (pt1) at (-1,0.5) {};
         \node[rectangle, draw, fill=black, minimum width=5 pt, minimum height=5 pt,inner sep=0pt] (pt2) at (-1,-0.5) {};
        \node[fill=black, circle, minimum size=5pt, inner sep=0pt, line width=1pt, draw=black] at (1,0) {};
        \node[fill=black, circle, minimum size=5pt, inner sep=0pt, line width=1pt, draw=black] at (0,0) {};
\end{tikzpicture}.
\label{eq:two_loop_deco}
\end{align}
We refer to this procedure as performing a one-loop chamber decomposition with respect to $y_1$. Remarkably, in \cite{Ferro:2024vwn} it was found that the terms appearing on the right hand side factorise as
\begin{align}
\begin{tikzpicture}[baseline=-0.5ex,scale=0.8]
        \draw[ultra thick,green] (0,0) -- (1,0);
        \draw[ultra thick,green]  (-1,0.5)--(0,0);
        \draw[ultra thick,red]  (-1,-0.5)--(0,0);
        \node[left=4pt] at (-1,0.5) {$\vertsGreen{\alpha}$};
        \node[left=4pt] at (-1,-0.5) {$\vertsRed{\alpha}$};
         \node[below=4pt] at (0,0) {$y_1$};
         \node[below=4pt] at (1,0) {$y_2$};
         \node[rectangle, draw, fill=black, minimum width=5 pt, minimum height=5 pt,inner sep=0pt] (pt1) at (-1,0.5) {};
         \node[rectangle, draw, fill=black, minimum width=5 pt, minimum height=5 pt,inner sep=0pt] (pt2) at (-1,-0.5) {};
        \node[fill=black, circle, minimum size=5pt, inner sep=0pt, line width=1pt, draw=black] at (1,0) {};
        \node[fill=black, circle, minimum size=5pt, inner sep=0pt, line width=1pt, draw=black] at (0,0) {};
\end{tikzpicture} & = \begin{tikzpicture}[baseline=-0.5ex,scale=0.8]
        \draw[ultra thick,green]  (-1,0.5)--(-2,0);
        \draw[ultra thick,red]  (-1,-0.5)--(-2,0);
        \node[right=4pt] at (-1,0.5) {$\vertsGreen{\alpha}$};
        \node[right=4pt] at (-1,-0.5) {$\vertsRed{\alpha}$};
         \node[below=4pt] at (-2,0) {$y_1$};
         \node[rectangle, draw, fill=black, minimum width=5 pt, minimum height=5 pt,inner sep=0pt] (pt1) at (-1,0.5) {};
         \node[rectangle, draw, fill=black, minimum width=5 pt, minimum height=5 pt,inner sep=0pt] (pt2) at (-1,-0.5) {};
        \node[fill=black, circle, minimum size=5pt, inner sep=0pt, line width=1pt, draw=black] at (-2,0) {};
\end{tikzpicture} \wedge  \begin{tikzpicture}[baseline=-0.5ex,scale=0.8]
        \draw[ultra thick,green] (0,0) -- (1,0);
        \draw[ultra thick,green]  (-1,0.5)--(0,0);
        \draw[ultra thick,red]  (-1,-0.5)--(0,0);
        \node[left=4pt] at (-1,0.5) {$\vertsGreen{\alpha}$};
        \node[left=4pt] at (-1,-0.5) {$\vertsRed{\alpha}$};
         \node[below=4pt] at (0,0) {$y_1$};
         \node[below=4pt] at (1,0) {$y_2$};
         \node[rectangle, draw, fill=black, minimum width=5 pt, minimum height=5 pt,inner sep=0pt] (pt1) at (-1,0.5) {};
         \node[rectangle, draw, fill=black, minimum width=5 pt, minimum height=5 pt,inner sep=0pt] (pt2) at (-1,-0.5) {};
        \node[fill=black, circle, minimum size=5pt, inner sep=0pt, line width=1pt, draw=black] at (1,0) {};
        \node[rectangle, draw, fill=black, minimum width=5 pt, minimum height=5 pt,inner sep=0pt] at (0,0) {};
\end{tikzpicture},
\end{align}
where the first factor is the canonical form for the one-loop chamber $\Omega_{y_1}\left[\mathfrak{c}_\alpha\right]$, and the second factor, which we refer to as the positive two-loop fiber, is a form in $y_2$ with $y_1$ treated as a fixed point inside the chamber $\mathfrak{c}_\alpha$ as indicated by the frozen vertex. In \cite{Ferro:2024vwn} explicit expressions for the canonical forms of the positive two-loop fibers were provided as a sum over chiral box integrands as
\begin{align}
 \begin{tikzpicture}[baseline=-0.5ex,scale=0.8]
        \draw[ultra thick,green] (0,0) -- (1,0);
        \draw[ultra thick,green]  (-1,0.5)--(0,0);
        \draw[ultra thick,red]  (-1,-0.5)--(0,0);
        \node[left=4pt] at (-1,0.5) {$\vertsGreen{\alpha}$};
        \node[left=4pt] at (-1,-0.5) {$\vertsRed{\alpha}$};
         \node[below=4pt] at (0,0) {$y_1$};
         \node[below=4pt] at (1,0) {$y_2$};
         \node[rectangle, draw, fill=black, minimum width=5 pt, minimum height=5 pt,inner sep=0pt] (pt1) at (-1,0.5) {};
         \node[rectangle, draw, fill=black, minimum width=5 pt, minimum height=5 pt,inner sep=0pt] (pt2) at (-1,-0.5) {};
        \node[fill=black, circle, minimum size=5pt, inner sep=0pt, line width=1pt, draw=black] at (1,0) {};
        \node[rectangle, draw, fill=black, minimum width=5 pt, minimum height=5 pt,inner sep=0pt] at (0,0) {};
\end{tikzpicture} =  \sum_{\ell^*_{ij} \in \vertsGreen{\alpha} }  \begin{tikzpicture}[baseline=-0.5ex,scale=0.8]
        \draw[ultra thick,black] (-1,0) -- (0,0) -- (1,0);
        \node[below=4pt] at (-1,0) {$\ell^*_{ij}$};
         \node[below=4pt] at (0,0) {$y_1$};
         \node[below=4pt] at (1,0) {$y_2$};
          \node[rectangle, draw, fill=black, minimum width=5 pt, minimum height=5 pt,inner sep=0pt] (pt1) at (-1,0) {};
        \node[rectangle, draw, fill=black, minimum width=5 pt, minimum height=5 pt,inner sep=0pt] (A) at (0,0) {};
        \node[fill=black, circle, minimum size=5pt, inner sep=0pt, line width=1pt, draw=black] (B) at (1,0) {};
\end{tikzpicture}.
\label{eq:sum_fiber_cb}
\end{align}
Substituting these results into \eqref{eq:two_loop_deco} and using \eqref{eq:sum_trick} the two-loop positive ladder can be written as
\begin{align}
\begin{tikzpicture}[baseline=-0.5ex,scale=0.8]
        \draw[ultra thick,green] (0,0) -- (1,0);
         \node[below=4pt] at (0,0) {$y_1$};
         \node[below=4pt] at (1,0) {$y_2$};
        \node[fill=black, circle, minimum size=5pt, inner sep=0pt, line width=1pt, draw=black] (A) at (0,0) {};
        \node[fill=black, circle, minimum size=5pt, inner sep=0pt, line width=1pt, draw=black] (B) at (1,0) {};
\end{tikzpicture} &  = \sum_{\ell^*_{ij} \in \vertsGreen{}} \begin{tikzpicture}[baseline=-0.5ex,scale=0.8]
        \draw[ultra thick,green] (0,0) -- (1,0);
         \node[below=4pt] at (0,0) {$y_1$};
         \node[below=4pt] at (1,0) {$\ell^*_{ij}$};
        \node[fill=black, circle, minimum size=5pt, inner sep=0pt, line width=1pt, draw=black] (A) at (0,0) {};
        \node[rectangle, draw, fill=black, minimum width=5 pt, minimum height=5 pt,inner sep=0pt] (B) at (1,0) {};
\end{tikzpicture} \wedge \begin{tikzpicture}[baseline=-0.5ex,scale=0.8]
        \draw[ultra thick,black] (-1,0) -- (0,0) -- (1,0);
        \node[below=4pt] at (-1,0) {$\ell^*_{ij}$};
         \node[below=4pt] at (0,0) {$y_1$};
         \node[below=4pt] at (1,0) {$y_2$};
          \node[rectangle, draw, fill=black, minimum width=5 pt, minimum height=5 pt,inner sep=0pt] (pt1) at (-1,0) {};
        \node[rectangle, draw, fill=black, minimum width=5 pt, minimum height=5 pt,inner sep=0pt] (A) at (0,0) {};
        \node[fill=black, circle, minimum size=5pt, inner sep=0pt, line width=1pt, draw=black] (B) at (1,0) {};
\end{tikzpicture},
\label{eq:two_loop_pos}
\end{align}
where we have used the fact that
\begin{align}
\begin{tikzpicture}[baseline=-0.5ex,scale=0.8]
        \draw[ultra thick,green] (0,0) -- (1,0);
         \node[below=4pt] at (0,0) {$y_1$};
         \node[below=3pt] at (1,0) {$\ell^*_{ij}$};
        \node[fill=black, circle, minimum size=5pt, inner sep=0pt, line width=1pt, draw=black] (A) at (0,0) {};
        \node[rectangle, draw, fill=black, minimum width=5 pt, minimum height=5 pt,inner sep=0pt] (B) at (1,0) {};
\end{tikzpicture}=\sum_{\alpha \in \mathfrak{C}_{ij}^+} \begin{tikzpicture}[baseline=-0.5ex,scale=0.8]
        \draw[ultra thick,green]  (-1,0.5)--(-2,0);
        \draw[ultra thick,red]  (-1,-0.5)--(-2,0);
        \node[right=4pt] at (-1,0.5) {$\vertsGreen{\alpha}$};
        \node[right=4pt] at (-1,-0.5) {$\vertsRed{\alpha}$};
         \node[below=4pt] at (-2,0) {$y_1$};
         \node[rectangle, draw, fill=black, minimum width=5 pt, minimum height=5 pt,inner sep=0pt] (pt1) at (-1,0.5) {};
         \node[rectangle, draw, fill=black, minimum width=5 pt, minimum height=5 pt,inner sep=0pt] (pt2) at (-1,-0.5) {};
        \node[fill=black, circle, minimum size=5pt, inner sep=0pt, line width=1pt, draw=black] at (-2,0) {};
\end{tikzpicture}.
\end{align}
As shown in \cite{Ferro:2024vwn} the canonical forms for these regions have simple expressions as sums of lower-point one-loop fibers
\begin{align}
\begin{tikzpicture}[baseline=-0.5ex,scale=0.8]
        \draw[ultra thick,green] (0,0) -- (1,0);
         \node[below=4pt] at (0,0) {$y_1$};
         \node[below=4pt] at (1,0) {$\ell^*_{ij}$};
        \node[fill=black, circle, minimum size=5pt, inner sep=0pt, line width=1pt, draw=black] (A) at (0,0) {};
        \node[rectangle, draw, fill=black, minimum width=5 pt, minimum height=5 pt,inner sep=0pt] (B) at (1,0) {};
\end{tikzpicture} =\Omega_{y_1}\left[\Delta({\bf x}_{ij})\right]+\Omega_{y_1}\left[\Delta({\bf x}_{ji})\right],
\end{align}
where we have defined $\Omega_{y_1}\left[\Delta({\bf x}_{ij})\right]$ and $\Omega_{y_1}\left[\Delta({\bf x}_{ij})\right]$ as the canonical forms for the one-loop fiber evaluated on points ${\bf x}_{ij}$ and ${\bf x}_{ji}$ respectively with
\begin{align}
{\bf x}_{ij}&= \{ x_{i+1},\ldots,x_j,\ell^*_{ij} \},
&&{\bf x}_{ji}= \{ x_{j+1},\ldots,x_i,\ell^*_{ij} \}.
\end{align}
This argument can be repeated for the two-loop negative ladder. We begin by performing a chamber decomposition with respect to $y_1$ which again factorises term-wise as 
\begin{align}
\begin{tikzpicture}[baseline=-0.5ex,scale=0.8]
        \draw[ultra thick,red] (0,0) -- (1,0);
         \node[below=4pt] at (0,0) {$y_1$};
         \node[below=4pt] at (1,0) {$y_2$};
        \node[fill=black, circle, minimum size=5pt, inner sep=0pt, line width=1pt, draw=black] (A) at (0,0) {};
        \node[fill=black, circle, minimum size=5pt, inner sep=0pt, line width=1pt, draw=black] (B) at (1,0) {};
\end{tikzpicture} & =\sum_{\alpha \in [n_{\mathfrak{C}}]} \begin{tikzpicture}[baseline=-0.5ex,scale=0.8]
        \draw[ultra thick,green]  (-1,0.5)--(-2,0);
        \draw[ultra thick,red]  (-1,-0.5)--(-2,0);
        \node[right=4pt] at (-1,0.5) {$\vertsGreen{\alpha}$};
        \node[right=4pt] at (-1,-0.5) {$\vertsRed{\alpha}$};
         \node[below=4pt] at (-2,0) {$y_1$};
         \node[rectangle, draw, fill=black, minimum width=5 pt, minimum height=5 pt,inner sep=0pt] (pt1) at (-1,0.5) {};
         \node[rectangle, draw, fill=black, minimum width=5 pt, minimum height=5 pt,inner sep=0pt] (pt2) at (-1,-0.5) {};
        \node[fill=black, circle, minimum size=5pt, inner sep=0pt, line width=1pt, draw=black] at (-2,0) {};
\end{tikzpicture} \wedge  \begin{tikzpicture}[baseline=-0.5ex,scale=0.8]
        \draw[ultra thick,red] (0,0) -- (1,0);
        \draw[ultra thick,green]  (-1,0.5)--(0,0);
        \draw[ultra thick,red]  (-1,-0.5)--(0,0);
        \node[left=4pt] at (-1,0.5) {$\vertsGreen{\alpha}$};
        \node[left=4pt] at (-1,-0.5) {$\vertsRed{\alpha}$};
         \node[below=4pt] at (0,0) {$y_1$};
         \node[below=4pt] at (1,0) {$y_2$};
         \node[rectangle, draw, fill=black, minimum width=5 pt, minimum height=5 pt,inner sep=0pt] (pt1) at (-1,0.5) {};
         \node[rectangle, draw, fill=black, minimum width=5 pt, minimum height=5 pt,inner sep=0pt] (pt2) at (-1,-0.5) {};
        \node[fill=black, circle, minimum size=5pt, inner sep=0pt, line width=1pt, draw=black] at (1,0) {};
        \node[rectangle, draw, fill=black, minimum width=5 pt, minimum height=5 pt,inner sep=0pt] at (0,0) {};
\end{tikzpicture}.
\label{eq:two_loop_deco_neg}
\end{align}
The canonical forms for the negative two-loop fibers can also be expressed as a sum over chiral box integrands as
\begin{align}
 \begin{tikzpicture}[baseline=-0.5ex,scale=0.8]
        \draw[ultra thick,red] (0,0) -- (1,0);
        \draw[ultra thick,green]  (-1,0.5)--(0,0);
        \draw[ultra thick,red]  (-1,-0.5)--(0,0);
        \node[left=4pt] at (-1,0.5) {$\vertsGreen{\alpha}$};
        \node[left=4pt] at (-1,-0.5) {$\vertsRed{\alpha}$};
         \node[below=4pt] at (0,0) {$y_1$};
         \node[below=4pt] at (1,0) {$y_2$};
         \node[rectangle, draw, fill=black, minimum width=5 pt, minimum height=5 pt,inner sep=0pt] (pt1) at (-1,0.5) {};
         \node[rectangle, draw, fill=black, minimum width=5 pt, minimum height=5 pt,inner sep=0pt] (pt2) at (-1,-0.5) {};
        \node[fill=black, circle, minimum size=5pt, inner sep=0pt, line width=1pt, draw=black] at (1,0) {};
        \node[rectangle, draw, fill=black, minimum width=5 pt, minimum height=5 pt,inner sep=0pt] at (0,0) {};
\end{tikzpicture} =  \sum_{\ell^*_{ij} \in \vertsRed{\alpha} }  \begin{tikzpicture}[baseline=-0.5ex,scale=0.8]
        \draw[ultra thick,black] (-1,0) -- (0,0) -- (1,0);
        \node[below=4pt] at (-1,0) {$\ell^*_{ij}$};
         \node[below=4pt] at (0,0) {$y_1$};
         \node[below=4pt] at (1,0) {$y_2$};
          \node[rectangle, draw, fill=black, minimum width=5 pt, minimum height=5 pt,inner sep=0pt] (pt1) at (-1,0) {};
        \node[rectangle, draw, fill=black, minimum width=5 pt, minimum height=5 pt,inner sep=0pt] (A) at (0,0) {};
        \node[fill=black, circle, minimum size=5pt, inner sep=0pt, line width=1pt, draw=black] (B) at (1,0) {};
\end{tikzpicture},
\label{eq:expand_cc_cb}
\end{align}
which when substituted into \eqref{eq:two_loop_deco_neg} results in the following form for the two-loop negative ladder
\begin{align}
\begin{tikzpicture}[baseline=-0.5ex,scale=0.8]
        \draw[ultra thick,red] (0,0) -- (1,0);
         \node[below=4pt] at (0,0) {$y_1$};
         \node[below=4pt] at (1,0) {$y_2$};
        \node[fill=black, circle, minimum size=5pt, inner sep=0pt, line width=1pt, draw=black] (A) at (0,0) {};
        \node[fill=black, circle, minimum size=5pt, inner sep=0pt, line width=1pt, draw=black] (B) at (1,0) {};
\end{tikzpicture} &  = \sum_{\ell^*_{ij} \in \vertsGreen{\alpha}} \begin{tikzpicture}[baseline=-0.5ex,scale=0.8]
        \draw[ultra thick,red] (0,0) -- (1,0);
         \node[below=4pt] at (0,0) {$y_1$};
         \node[below=4pt] at (1,0) {$\ell^*_{ij}$};
        \node[fill=black, circle, minimum size=5pt, inner sep=0pt, line width=1pt, draw=black] (A) at (0,0) {};
        \node[rectangle, draw, fill=black, minimum width=5 pt, minimum height=5 pt,inner sep=0pt] (B) at (1,0) {};
\end{tikzpicture} \wedge \begin{tikzpicture}[baseline=-0.5ex,scale=0.8]
        \draw[ultra thick,black] (-1,0) -- (0,0) -- (1,0);
        \node[below=4pt] at (-1,0) {$\ell^*_{ij}$};
         \node[below=4pt] at (0,0) {$y_1$};
         \node[below=4pt] at (1,0) {$y_2$};
          \node[rectangle, draw, fill=black, minimum width=5 pt, minimum height=5 pt,inner sep=0pt] (pt1) at (-1,0) {};
        \node[rectangle, draw, fill=black, minimum width=5 pt, minimum height=5 pt,inner sep=0pt] (A) at (0,0) {};
        \node[fill=black, circle, minimum size=5pt, inner sep=0pt, line width=1pt, draw=black] (B) at (1,0) {};
\end{tikzpicture},
\label{eq:two_loop_neg}
\end{align}
where
\begin{align}
\begin{tikzpicture}[baseline=-0.5ex,scale=0.8]
        \draw[ultra thick,red] (0,0) -- (1,0);
         \node[below=4pt] at (0,0) {$y_1$};
         \node[below=4pt] at (1,0) {$\ell^*_{ij}$};
        \node[fill=black, circle, minimum size=5pt, inner sep=0pt, line width=1pt, draw=black] (A) at (0,0) {};
        \node[rectangle, draw, fill=black, minimum width=5 pt, minimum height=5 pt,inner sep=0pt] (B) at (1,0) {};
\end{tikzpicture} = \Omega_{y_1}[\Delta({\bf x})]- \Omega_{y_1}[\Delta({\bf x}_{ij})]-\Omega_{y_1}[\Delta({\bf x}_{ji})],
\label{eq:subamps_neg}
\end{align}
is again written in terms of canonical forms of one-loop fibers. 

With the simplification \eqref{eq:simplification_n4} for $n=4$, the canonical form for the one-loop fiber can be factored out of the two-loop positive/negative ladders as
\begin{align}
 \begin{tikzpicture}[baseline=-0.5ex,scale=0.8]
         \node[below=4pt] at (0,0) {$y_1$};
         \node[below=4pt] at (1,0) {$y_2$};
         \draw[ultra thick,green]  (0,0)--(1,0);
        \node[fill=black, circle, minimum size=5pt, inner sep=0pt, line width=1pt, draw=black] at (0,0) {};
         \node[fill=black, circle, minimum size=5pt, inner sep=0pt, line width=1pt, draw=black] at (1,0) {};
\end{tikzpicture} &= \begin{tikzpicture}[baseline=-0.5ex,scale=0.8]
         \node[below=4pt] at (0,0) {$y_1$};
        \node[fill=black, circle, minimum size=5pt, inner sep=0pt, line width=1pt, draw=black] (A) at (0,0) {};
\end{tikzpicture} \wedge \sum_{x_i \in \mathcal{V}^+} \begin{tikzpicture}[baseline=-0.5ex,scale=0.8]
        \draw[ultra thick,black]  (-3,0)--(-2,0);
        \node[below=4pt] at (-3,0) {{\small $x_i$}};
        \node[below=4pt] at (-2,0) {{\small $y_1$}};
        \node[below=4pt] at (-1,0) {$y_2$};
         \draw[ultra thick,black]  (-2,0)--(-1,0);
         \node[rectangle, draw, fill=black, minimum width=5 pt, minimum height=5 pt,inner sep=0pt] (pt1) at (-3,0) {};
        \node[fill=black, circle, minimum size=5pt, inner sep=0pt, line width=1pt, draw=black] at (-1,0) {};
        \node[rectangle, draw, fill=black, minimum width=5 pt, minimum height=5 pt,inner sep=0pt] (pt2) at (-2,0) {};
\end{tikzpicture}, \notag \\
\begin{tikzpicture}[baseline=-0.5ex,scale=0.8]
         \node[below=4pt] at (0,0) {$y_1$};
         \node[below=4pt] at (1,0) {$y_2$};
         \draw[ultra thick,red]  (0,0)--(1,0);
        \node[fill=black, circle, minimum size=5pt, inner sep=0pt, line width=1pt, draw=black] at (0,0) {};
         \node[fill=black, circle, minimum size=5pt, inner sep=0pt, line width=1pt, draw=black] at (1,0) {};
\end{tikzpicture} &= \begin{tikzpicture}[baseline=-0.5ex,scale=0.8]
         \node[below=4pt] at (0,0) {$y_1$};
        \node[fill=black, circle, minimum size=5pt, inner sep=0pt, line width=1pt, draw=black] (A) at (0,0) {};
\end{tikzpicture} \wedge \sum_{\ell^*_{ij} \in \mathcal{V}^-} \begin{tikzpicture}[baseline=-0.5ex,scale=0.8]
        \draw[ultra thick,black]  (-3,0)--(-2,0);
        \node[below=4pt] at (-3,0) {{\small $\ell^*_{ij}$}};
        \node[below=4pt] at (-2,0) {{\small $y_1$}};
        \node[below=4pt] at (-1,0) {$y_2$};
         \draw[ultra thick,black]  (-2,0)--(-1,0);
         \node[rectangle, draw, fill=black, minimum width=5 pt, minimum height=5 pt,inner sep=0pt] (pt1) at (-3,0) {};
        \node[fill=black, circle, minimum size=5pt, inner sep=0pt, line width=1pt, draw=black] at (-1,0) {};
        \node[rectangle, draw, fill=black, minimum width=5 pt, minimum height=5 pt,inner sep=0pt] (pt2) at (-2,0) {};
\end{tikzpicture}.
\end{align}
No simplifications occur for $n>4$ and the two-loop positive and negative integrands are given by the general formulae \eqref{eq:two_loop_pos} and \eqref{eq:two_loop_neg}, respectively.
\section{General Ladders in Loop Space}
\label{sec:ladders}
We now move on to study how the two-loop positive and negative results of \eqref{eq:two_loop_pos} and \eqref{eq:two_loop_neg} generalise to arbitrary ladders in loop space. As a demonstrative example we consider the all negative ladder
\begin{align}
\begin{tikzpicture}[baseline=-0.5ex,scale=0.7]
        \draw[ultra thick,red] (0,0) -- (1,0)--(1.2,0);
        \draw[ultra thick,red] (1.8,0) -- (2,0) -- (3,0);
        \node[fill=black, circle, minimum size=4pt, inner sep=0pt, line width=1pt, draw=black] (A) at (0,0) {};
        \node[fill=black, circle, minimum size=4pt, inner sep=0pt, line width=1pt, draw=black] (B) at (1,0) {};
        \node[fill=black, circle, minimum size=4pt, inner sep=0pt, line width=1pt, draw=black] (C) at (2,0) {};
        \node[fill=black, circle, minimum size=4pt, inner sep=0pt, line width=1pt, draw=black] (D) at (3,0) {};
        \node[] at (1.55,0) {{\color{red} \small \ldots}};
         \node[below=4pt] at (0,0) {$y_1$};
         \node[below=4pt] at (1,0) {$y_2$};
         \node[below=4pt] at (3,0) {$y_{L+1}$};
\end{tikzpicture}  = \Omega_{y_1\ldots y_{L+1}}\left[\{ y_1,\ldots,y_{L+1} \in \Delta({\bf x}) : (y_{a+1}-y_a)^2<0 \text{ for all } a \in [L] \}\right]. \notag 
\end{align}
Our starting point will be to consider the $(L+1)$-loop negative ladder expanded over one-loop chambers with respect to $y_L$ which in analogy to the two-loop example can be written in the following term-wise factorised form 
\begin{align}
\begin{tikzpicture}[baseline=-0.5ex,scale=0.8]
        \draw[ultra thick,red] (0,0) -- (1,0)--(1.2,0);
        \draw[ultra thick,red] (1.8,0) -- (2,0) -- (3,0);
        \node[fill=black, circle, minimum size=4pt, inner sep=0pt, line width=1pt, draw=black] (A) at (0,0) {};
        \node[fill=black, circle, minimum size=4pt, inner sep=0pt, line width=1pt, draw=black] (B) at (1,0) {};
        \node[fill=black, circle, minimum size=4pt, inner sep=0pt, line width=1pt, draw=black] (C) at (2,0) {};
        \node[fill=black, circle, minimum size=4pt, inner sep=0pt, line width=1pt, draw=black] (D) at (3,0) {};
        \node[] at (1.55,0) {\textcolor{red}{{\small \ldots}}};
         \node[below=4pt] at (0,0) {$y_1$};
         \node[below=4pt] at (1,0) {$y_2$};
         \node[below=4pt] at (2,0) {$y_{L}$};
         \node[below=4pt] at (3,0) {$y_{L+1}$};
\end{tikzpicture}& =\sum_{\alpha \in [n_{\mathfrak{C}}] }\begin{tikzpicture}[baseline=-0.5ex,scale=0.8]
        \draw[ultra thick,red] (0,0) -- (1,0)--(1.2,0);
        \draw[ultra thick,red] (1.8,0) -- (2,0);
        \draw[ultra thick,green]  (3,0.5)--(2,0);
        \draw[ultra thick,red]  (3,-0.5)--(2,0);
        \node[fill=black, circle, minimum size=4pt, inner sep=0pt, line width=1pt, draw=black] (A) at (0,0) {};
        \node[fill=black, circle, minimum size=4pt, inner sep=0pt, line width=1pt, draw=black] (B) at (1,0) {};
        \node[fill=black, circle, minimum size=4pt, inner sep=0pt, line width=1pt, draw=black] (C) at (2,0) {};
         \node[rectangle, draw, fill=black, minimum width=5 pt, minimum height=5 pt,inner sep=0pt] (pt1) at (3,0.5) {};
         \node[rectangle, draw, fill=black, minimum width=5 pt, minimum height=5 pt,inner sep=0pt] (pt2) at (3,-0.5) {};
          \node[] at (1.55,0) {\textcolor{red}{{\small \ldots}}};
         \node[below=4pt] at (0,0) {$y_1$};
         \node[below=4pt] at (1,0) {$y_2$};
         \node[below=4pt] at (2,0) {$y_{L}$};
        \node[right=4pt] at (3,0.5) {$\vertsGreen{\alpha}$};
        \node[right=4pt] at (3,-0.5) {$\vertsRed{\alpha}$};
\end{tikzpicture}   \wedge \begin{tikzpicture}[baseline=-0.5ex,scale=0.8]
        \draw[ultra thick,red] (0,0) -- (1,0);
        \draw[ultra thick,green]  (-1,0.5)--(0,0);
        \draw[ultra thick,red]  (-1,-0.5)--(0,0);
        \node[left=4pt] at (-1,0.5) {$\vertsGreen{\alpha}$};
        \node[left=4pt] at (-1,-0.5) {$\vertsRed{\alpha}$};
         \node[below=4pt] at (0,0) {$y_L$};
         \node[below=4pt] at (1,0) {$y_{L+1}$};
         \node[rectangle, draw, fill=black, minimum width=5 pt, minimum height=5 pt,inner sep=0pt] (pt1) at (-1,0.5) {};
         \node[rectangle, draw, fill=black, minimum width=5 pt, minimum height=5 pt,inner sep=0pt] (pt2) at (-1,-0.5) {};
        \node[fill=black, circle, minimum size=5pt, inner sep=0pt, line width=1pt, draw=black] at (1,0) {};
        \node[rectangle, draw, fill=black, minimum width=5 pt, minimum height=5 pt,inner sep=0pt] at (0,0) {};
\end{tikzpicture}.
\label{eq:three_loop_ladder_int}
\end{align}
We notice the factor on the right is nothing other than the negative two-loop fiber which already appeared in the two-loop result \eqref{eq:two_loop_deco_neg}. Using the same steps as before, expanding the canonical forms for the negative two-loop fibers in terms of chiral boxes, and reorganising the sum, this can be written as
\begin{align}
\begin{tikzpicture}[baseline=-0.5ex,scale=0.8]
        \draw[ultra thick,red] (0,0) -- (1,0)--(1.2,0);
        \draw[ultra thick,red] (1.8,0) -- (2,0) -- (3,0);
        \node[fill=black, circle, minimum size=4pt, inner sep=0pt, line width=1pt, draw=black] (A) at (0,0) {};
        \node[fill=black, circle, minimum size=4pt, inner sep=0pt, line width=1pt, draw=black] (B) at (1,0) {};
        \node[fill=black, circle, minimum size=4pt, inner sep=0pt, line width=1pt, draw=black] (C) at (2,0) {};
        \node[fill=black, circle, minimum size=4pt, inner sep=0pt, line width=1pt, draw=black] (D) at (3,0) {};
        \node[] at (1.55,0) {\textcolor{red}{{\small \ldots}}};
         \node[below=4pt] at (0,0) {$y_1$};
         \node[below=4pt] at (1,0) {$y_2$};
         \node[below=4pt] at (2,0) {$y_{L}$};
         \node[below=4pt] at (3,0) {$y_{L+1}$};
\end{tikzpicture}& =\sum_{\ell^*_{ij} \in \mathcal{V}^-}\begin{tikzpicture}[baseline=-0.5ex,scale=0.8]
        \draw[ultra thick,red] (0,0) -- (1,0)--(1.2,0);
        \draw[ultra thick,red] (1.8,0) -- (2,0);
        \draw[ultra thick,red]  (3,0)--(2,0);
        \node[fill=black, circle, minimum size=4pt, inner sep=0pt, line width=1pt, draw=black] (A) at (0,0) {};
        \node[fill=black, circle, minimum size=4pt, inner sep=0pt, line width=1pt, draw=black] (B) at (1,0) {};
        \node[fill=black, circle, minimum size=4pt, inner sep=0pt, line width=1pt, draw=black] (C) at (2,0) {};
         \node[rectangle, draw, fill=black, minimum width=5 pt, minimum height=5 pt,inner sep=0pt] (pt1) at (3,0) {};
          \node[] at (1.55,0) {\textcolor{red}{{\small \ldots}}};
         \node[below=4pt] at (0,0) {$y_1$};
         \node[below=4pt] at (1,0) {$y_2$};
         \node[below=4pt] at (2,0) {$y_{L}$};
        \node[below=4pt] at (3,0) {$\ell^*_{ij}$};
\end{tikzpicture}   \wedge \begin{tikzpicture}[baseline=-0.5ex,scale=0.8]
        \draw[ultra thick,black] (0,0) -- (1,0);
        \draw[ultra thick,black]  (-1,0)--(0,0);
        \node[below=4pt] at (-1,0) {$\ell^*_{ij}$};
         \node[below=4pt] at (0,0) {$y_L$};
         \node[below=4pt] at (1,0) {$y_{L+1}$};
         \node[rectangle, draw, fill=black, minimum width=5 pt, minimum height=5 pt,inner sep=0pt] (pt1) at (-1,0) {};
        \node[fill=black, circle, minimum size=5pt, inner sep=0pt, line width=1pt, draw=black] at (1,0) {};
        \node[rectangle, draw, fill=black, minimum width=5 pt, minimum height=5 pt,inner sep=0pt] at (0,0) {};
\end{tikzpicture}.
\label{eq:ladder_int}
\end{align}
Next we consider the coefficients multiplying each chiral box expanded in terms of one-loop chambers, this time with respect to $y_{L-1}$, which can be written as 
\begin{align}
\begin{tikzpicture}[baseline=-0.5ex,scale=0.8]
        \draw[ultra thick,red] (0,0) -- (1,0)--(1.2,0);
        \draw[ultra thick,red] (1.8,0) -- (2,0) -- (3,0);
        \node[fill=black, circle, minimum size=4pt, inner sep=0pt, line width=1pt, draw=black] (A) at (0,0) {};
        \node[fill=black, circle, minimum size=4pt, inner sep=0pt, line width=1pt, draw=black] (B) at (1,0) {};
        \node[fill=black, circle, minimum size=4pt, inner sep=0pt, line width=1pt, draw=black] (C) at (2,0) {};
        \node[rectangle, draw, fill=black, minimum width=5 pt, minimum height=5 pt,inner sep=0pt] (pt1) at (3,0) {};
        \node[] at (1.55,0) {\textcolor{red}{{\small \ldots}}};
         \node[below=4pt] at (0,0) {$y_1$};
         \node[below=4pt] at (1,0) {$y_2$};
         \node[below=4pt] at (2,0) {$y_{L}$};
         \node[below=4pt] at (3,0) {$\ell^*_{ij}$};
\end{tikzpicture}& =\sum_{\alpha \in [n_{\mathfrak{C}}]}\begin{tikzpicture}[baseline=-0.5ex,scale=0.8]
        \draw[ultra thick,red] (0,0) -- (1,0)--(1.2,0);
        \draw[ultra thick,red] (1.8,0) -- (2,0);
        \draw[ultra thick,green]  (3,0.5)--(2,0);
        \draw[ultra thick,red]  (3,-0.5)--(2,0);
        \node[fill=black, circle, minimum size=4pt, inner sep=0pt, line width=1pt, draw=black] (A) at (0,0) {};
        \node[fill=black, circle, minimum size=4pt, inner sep=0pt, line width=1pt, draw=black] (B) at (1,0) {};
        \node[fill=black, circle, minimum size=4pt, inner sep=0pt, line width=1pt, draw=black] (C) at (2,0) {};
         \node[rectangle, draw, fill=black, minimum width=5 pt, minimum height=5 pt,inner sep=0pt] (pt1) at (3,0.5) {};
         \node[rectangle, draw, fill=black, minimum width=5 pt, minimum height=5 pt,inner sep=0pt] (pt2) at (3,-0.5) {};
          \node[] at (1.55,0) {\textcolor{red}{{\small \ldots}}};
         \node[below=4pt] at (0,0) {$y_1$};
         \node[below=4pt] at (1,0) {$y_2$};
         \node[below=4pt] at (2,0) {$y_{L-1}$};
        \node[right=4pt] at (3,0.5) {$\vertsGreen{\alpha}$};
        \node[right=4pt] at (3,-0.5) {$\vertsRed{\alpha}$};
\end{tikzpicture}   \wedge  \begin{tikzpicture}[baseline=-0.5ex,scale=0.8]
        \draw[ultra thick,red] (0,0) -- (2,0);
        \draw[ultra thick,green]  (-1,0.5)--(0,0);
        \draw[ultra thick,red]  (-1,-0.5)--(0,0);
        \node[left=4pt] at (-1,0.5) {$\vertsGreen{\alpha}$};
        \node[left=4pt] at (-1,-0.5) {$\vertsRed{\alpha}$};
         \node[below=4pt] at (0,0) {$y_{L-1}$};
         \node[below=4pt] at (1,0) {$y_L$};
         \node[below=4pt] at (2,0) {$\ell^*_{ij}$};
         \node[rectangle, draw, fill=black, minimum width=5 pt, minimum height=5 pt,inner sep=0pt] (pt1) at (-1,0.5) {};
         \node[rectangle, draw, fill=black, minimum width=5 pt, minimum height=5 pt,inner sep=0pt] (pt2) at (-1,-0.5) {};
        \node[fill=black, circle, minimum size=5pt, inner sep=0pt, line width=1pt, draw=black] at (1,0) {};
        \node[rectangle, draw, fill=black, minimum width=5 pt, minimum height=5 pt,inner sep=0pt] at (0,0) {};
        \node[rectangle, draw, fill=black, minimum width=5 pt, minimum height=5 pt,inner sep=0pt] at (2,0) {};
\end{tikzpicture}.
\label{eq:box_coeffs}
\end{align}
The new factors appearing on the right hand side are the canonical forms of the subset of the negative two-loop fiber that satisfies the additional constraint $(y_2-\ell^*_{ij})^2<0$. Along with its $(y_2-\ell^*_{ij})^2>0$ counterpart these two objects decompose each negative two-loop fiber into two regions as  
\begin{align}
\begin{tikzpicture}[baseline=-0.5ex,scale=0.8]
        \draw[ultra thick,red] (0,0) -- (1,0);
        \draw[ultra thick,green]  (-1,0.5)--(0,0);
        \draw[ultra thick,red]  (-1,-0.5)--(0,0);
        \node[left=4pt] at (-1,0.5) {$\vertsGreen{\alpha}$};
        \node[left=4pt] at (-1,-0.5) {$\vertsRed{\alpha}$};
         \node[below=4pt] at (0,0) {$y_{L-1}$};
         \node[below=4pt] at (1,0) {$y_L$};
         \node[rectangle, draw, fill=black, minimum width=5 pt, minimum height=5 pt,inner sep=0pt] (pt1) at (-1,0.5) {};
         \node[rectangle, draw, fill=black, minimum width=5 pt, minimum height=5 pt,inner sep=0pt] (pt2) at (-1,-0.5) {};
        \node[fill=black, circle, minimum size=5pt, inner sep=0pt, line width=1pt, draw=black] at (1,0) {};
        \node[rectangle, draw, fill=black, minimum width=5 pt, minimum height=5 pt,inner sep=0pt] at (0,0) {};
\end{tikzpicture}=\begin{tikzpicture}[baseline=-0.5ex,scale=0.8]
        \draw[ultra thick,red] (0,0) -- (1,0);
        \draw[ultra thick,green] (1,0) -- (2,0);
        \draw[ultra thick,green]  (-1,0.5)--(0,0);
        \draw[ultra thick,red]  (-1,-0.5)--(0,0);
        \node[left=4pt] at (-1,0.5) {$\vertsGreen{\alpha}$};
        \node[left=4pt] at (-1,-0.5) {$\vertsRed{\alpha}$};
         \node[below=4pt] at (0,0) {$y_{L-1}$};
         \node[below=4pt] at (1,0) {$y_L$};
         \node[below=4pt] at (2,0) {$\ell^*_{ij}$};
         \node[rectangle, draw, fill=black, minimum width=5 pt, minimum height=5 pt,inner sep=0pt] (pt1) at (-1,0.5) {};
         \node[rectangle, draw, fill=black, minimum width=5 pt, minimum height=5 pt,inner sep=0pt] (pt2) at (-1,-0.5) {};
        \node[fill=black, circle, minimum size=5pt, inner sep=0pt, line width=1pt, draw=black] at (1,0) {};
        \node[rectangle, draw, fill=black, minimum width=5 pt, minimum height=5 pt,inner sep=0pt] at (0,0) {};
        \node[rectangle, draw, fill=black, minimum width=5 pt, minimum height=5 pt,inner sep=0pt] at (2,0) {};
\end{tikzpicture}+\begin{tikzpicture}[baseline=-0.5ex,scale=0.8]
        \draw[ultra thick,red] (0,0) -- (2,0);
        \draw[ultra thick,green]  (-1,0.5)--(0,0);
        \draw[ultra thick,red]  (-1,-0.5)--(0,0);
        \node[left=4pt] at (-1,0.5) {$\vertsGreen{\alpha}$};
        \node[left=4pt] at (-1,-0.5) {$\vertsRed{\alpha}$};
         \node[below=4pt] at (0,0) {$y_{L-1}$};
         \node[below=4pt] at (1,0) {$y_L$};
         \node[below=4pt] at (2,0) {$\ell^*_{ij}$};
         \node[rectangle, draw, fill=black, minimum width=5 pt, minimum height=5 pt,inner sep=0pt] (pt1) at (-1,0.5) {};
         \node[rectangle, draw, fill=black, minimum width=5 pt, minimum height=5 pt,inner sep=0pt] (pt2) at (-1,-0.5) {};
        \node[fill=black, circle, minimum size=5pt, inner sep=0pt, line width=1pt, draw=black] at (1,0) {};
        \node[rectangle, draw, fill=black, minimum width=5 pt, minimum height=5 pt,inner sep=0pt] at (0,0) {};
        \node[rectangle, draw, fill=black, minimum width=5 pt, minimum height=5 pt,inner sep=0pt] at (2,0) {};
\end{tikzpicture}.
\label{eq:decomposetwoloopfiber}
\end{align}
To continue we now wish to perform an expansion of the terms appearing on the right hand side of \eqref{eq:decomposetwoloopfiber}, analogous to the expansion \eqref{eq:sum_fiber_cb} of the positive/negative two-loop fibers in terms of chiral boxes. While it is not obvious that such an expansion exists, remarkably, we find that it can indeed be constructed and takes a simple form
\begin{align}
 \begin{tikzpicture}[baseline=-0.5ex,scale=0.8]
        \draw[ultra thick,red] (0,0) -- (1,0);
        \draw[ultra thick,green] (1,0) -- (2,0);
        \draw[ultra thick,green]  (-1,0.5)--(0,0);
        \draw[ultra thick,red]  (-1,-0.5)--(0,0);
        \node[left=4pt] at (-1,0.5) {$\vertsGreen{\alpha}$};
        \node[left=4pt] at (-1,-0.5) {$\vertsRed{\alpha}$};
         \node[below=4pt] at (0,0) {$y_{L-1}$};
         \node[below=4pt] at (1,0) {$y_L$};
          \node[below=4pt] at (2,0) {$\ell^*_{ij}$};
         \node[rectangle, draw, fill=black, minimum width=5 pt, minimum height=5 pt,inner sep=0pt] (pt1) at (-1,0.5) {};
         \node[rectangle, draw, fill=black, minimum width=5 pt, minimum height=5 pt,inner sep=0pt] (pt2) at (-1,-0.5) {};
        \node[fill=black, circle, minimum size=5pt, inner sep=0pt, line width=1pt, draw=black] at (1,0) {};
        \node[rectangle, draw, fill=black, minimum width=5 pt, minimum height=5 pt,inner sep=0pt] at (0,0) {};
        \node[rectangle, draw, fill=black, minimum width=5 pt, minimum height=5 pt,inner sep=0pt] at (2,0) {};
\end{tikzpicture} &=  \sum_{\ell^*_{kl} \in \vertsGreen{\alpha} }  \begin{tikzpicture}[baseline=-0.5ex,scale=0.8]
        \draw[ultra thick,black] (-1,0) -- (1,0);
        \draw[ultra thick,green]  (1,0) -- (2,0);
        \node[below=4pt] at (-1,0) {$\ell^*_{kl}$};
         \node[below=4pt] at (0,0) {$y_{L-1}$};
         \node[below=4pt] at (1,0) {$y_L$};
         \node[below=4pt] at (2,0) {$\ell^*_{ij}$};
          \node[rectangle, draw, fill=black, minimum width=5 pt, minimum height=5 pt,inner sep=0pt] (pt1) at (-1,0) {};
        \node[rectangle, draw, fill=black, minimum width=5 pt, minimum height=5 pt,inner sep=0pt] (A) at (0,0) {};
        \node[rectangle, draw, fill=black, minimum width=5 pt, minimum height=5 pt,inner sep=0pt] (A) at (2,0) {};
        \node[fill=black, circle, minimum size=5pt, inner sep=0pt, line width=1pt, draw=black] (B) at (1,0) {};
\end{tikzpicture}, \notag \\
 \begin{tikzpicture}[baseline=-0.5ex,scale=0.8]
        \draw[ultra thick,red] (0,0) -- (2,0);
        \draw[ultra thick,green]  (-1,0.5)--(0,0);
        \draw[ultra thick,red]  (-1,-0.5)--(0,0);
        \node[left=4pt] at (-1,0.5) {$\vertsGreen{\alpha}$};
        \node[left=4pt] at (-1,-0.5) {$\vertsRed{\alpha}$};
         \node[below=4pt] at (0,0) {$y_{L-1}$};
         \node[below=4pt] at (1,0) {$y_L$};
          \node[below=4pt] at (2,0) {$\ell^*_{ij}$};
         \node[rectangle, draw, fill=black, minimum width=5 pt, minimum height=5 pt,inner sep=0pt] (pt1) at (-1,0.5) {};
         \node[rectangle, draw, fill=black, minimum width=5 pt, minimum height=5 pt,inner sep=0pt] (pt2) at (-1,-0.5) {};
        \node[fill=black, circle, minimum size=5pt, inner sep=0pt, line width=1pt, draw=black] at (1,0) {};
        \node[rectangle, draw, fill=black, minimum width=5 pt, minimum height=5 pt,inner sep=0pt] at (0,0) {};
        \node[rectangle, draw, fill=black, minimum width=5 pt, minimum height=5 pt,inner sep=0pt] at (2,0) {};
\end{tikzpicture} &=  \sum_{\ell^*_{kl} \in \vertsRed{\alpha} }  \begin{tikzpicture}[baseline=-0.5ex,scale=0.8]
        \draw[ultra thick,black] (-1,0) -- (0,0) -- (1,0);
        \draw[ultra thick,red] (1,0) -- (2,0);
        \node[below=4pt] at (-1,0) {$\ell^*_{kl}$};
         \node[below=4pt] at (0,0) {$y_{L-1}$};
         \node[below=4pt] at (1,0) {$y_L$};
         \node[below=4pt] at (2,0) {$\ell^*_{ij}$};
          \node[rectangle, draw, fill=black, minimum width=5 pt, minimum height=5 pt,inner sep=0pt] (pt1) at (-1,0) {};
        \node[rectangle, draw, fill=black, minimum width=5 pt, minimum height=5 pt,inner sep=0pt] (A) at (0,0) {};
        \node[rectangle, draw, fill=black, minimum width=5 pt, minimum height=5 pt,inner sep=0pt] (A) at (2,0) {};
        \node[fill=black, circle, minimum size=5pt, inner sep=0pt, line width=1pt, draw=black] (B) at (1,0) {};
\end{tikzpicture}.
\end{align}
The terms appearing in the summation above split the chiral box integrand for point $\ell^*_{kl}$ into a positive and negative part with respect to $\ell^*_{ij}$ such that we have 
\begin{align}
\begin{tikzpicture}[baseline=-0.5ex,scale=0.8]
        \draw[ultra thick,black] (-1,0) -- (0,0) -- (1,0);
        \node[below=4pt] at (-1,0) {$\ell^*_{kl}$};
         \node[below=4pt] at (0,0) {$y_{L-1}$};
         \node[below=4pt] at (1,0) {$y_{L}$};
          \node[rectangle, draw, fill=black, minimum width=5 pt, minimum height=5 pt,inner sep=0pt] (pt1) at (-1,0) {};
        \node[rectangle, draw, fill=black, minimum width=5 pt, minimum height=5 pt,inner sep=0pt] (A) at (0,0) {};
        \node[fill=black, circle, minimum size=5pt, inner sep=0pt, line width=1pt, draw=black] (B) at (1,0) {};
\end{tikzpicture} = \begin{tikzpicture}[baseline=-0.5ex,scale=0.8]
        \draw[ultra thick,black] (0,0) -- (1,0);
        \draw[ultra thick,green] (1,0)--(2,0);
        \draw[ultra thick,black]  (-1,0)--(0,0);
        \node[below=4pt] at (2,0) {$\ell^*_{ij}$};
        \node[below=4pt] at (-1,0) {$\ell^*_{kl}$};
         \node[below=4pt] at (0,0) {$y_{L-1}$};
         \node[below=4pt] at (1,0) {$y_L$};
          \node[rectangle, draw, fill=black, minimum width=5 pt, minimum height=5 pt,inner sep=0pt] (pt1) at (-1,0) {};
        \node[rectangle, draw, fill=black, minimum width=5 pt, minimum height=5 pt,inner sep=0pt] (A) at (0,0) {};
        \node[fill=black, circle, minimum size=5pt, inner sep=0pt, line width=1pt, draw=black] (B) at (1,0) {};
        \node[rectangle, draw, fill=black, minimum width=5 pt, minimum height=5 pt,inner sep=0pt] at (2,0) {};
\end{tikzpicture}+\begin{tikzpicture}[baseline=-0.5ex,scale=0.8]
        \draw[ultra thick,black] (0,0) -- (1,0)--(2,0);
        \draw[ultra thick,red] (1,0)--(2,0);
        \draw[ultra thick,black]  (-1,0)--(0,0);
        \node[below=4pt] at (2,0) {$\ell^*_{ij}$};
        \node[below=4pt] at (-1,0) {$\ell^*_{kl}$};
         \node[below=4pt] at (0,0) {$y_{L-1}$};
         \node[below=4pt] at (1,0) {$y_L$};
          \node[rectangle, draw, fill=black, minimum width=5 pt, minimum height=5 pt,inner sep=0pt] (pt1) at (-1,0) {};
        \node[rectangle, draw, fill=black, minimum width=5 pt, minimum height=5 pt,inner sep=0pt] (A) at (0,0) {};
        \node[fill=black, circle, minimum size=5pt, inner sep=0pt, line width=1pt, draw=black] (B) at (1,0) {};
        \node[rectangle, draw, fill=black, minimum width=5 pt, minimum height=5 pt,inner sep=0pt] at (2,0) {};
\end{tikzpicture}.
\end{align}
There are five formulae depending on how $ij$ and $kl$ are distributed:
\begin{itemize}
\item $ij$ and $kl$ are crossing chords of the $n$-gon
\begin{align}
 \begin{tikzpicture}[baseline=-0.5ex,scale=0.8]
        \draw[ultra thick,black] (0,0) -- (1,0);
        \draw[ultra thick,green] (1,0)--(2,0);
        \draw[ultra thick,black]  (-1,0)--(0,0);
        \node[below=4pt] at (2,0) {$\ell^*_{kl}$};
        \node[below=4pt] at (-1,0) {$\ell^*_{ij}$};
         \node[below=4pt] at (0,0) {$y_a$};
         \node[below=4pt] at (1,0) {$y_b$};
          \node[rectangle, draw, fill=black, minimum width=5 pt, minimum height=5 pt,inner sep=0pt] (pt1) at (-1,0) {};
        \node[rectangle, draw, fill=black, minimum width=5 pt, minimum height=5 pt,inner sep=0pt] (A) at (0,0) {};
        \node[fill=black, circle, minimum size=5pt, inner sep=0pt, line width=1pt, draw=black] (B) at (1,0) {};
        \node[rectangle, draw, fill=black, minimum width=5 pt, minimum height=5 pt,inner sep=0pt] at (2,0) {};
\end{tikzpicture}&=0, \notag \\
 \begin{tikzpicture}[baseline=-0.5ex,scale=0.8]
        \draw[ultra thick,black] (0,0) -- (1,0);
        \draw[ultra thick,red] (1,0)--(2,0);
        \draw[ultra thick,black]  (-1,0)--(0,0);
        \node[below=4pt] at (2,0) {$\ell^*_{kl}$};
        \node[below=4pt] at (-1,0) {$\ell^*_{ij}$};
         \node[below=4pt] at (0,0) {$y_a$};
         \node[below=4pt] at (1,0) {$y_b$};
          \node[rectangle, draw, fill=black, minimum width=5 pt, minimum height=5 pt,inner sep=0pt] (pt1) at (-1,0) {};
        \node[rectangle, draw, fill=black, minimum width=5 pt, minimum height=5 pt,inner sep=0pt] (A) at (0,0) {};
        \node[fill=black, circle, minimum size=5pt, inner sep=0pt, line width=1pt, draw=black] (B) at (1,0) {};
        \node[rectangle, draw, fill=black, minimum width=5 pt, minimum height=5 pt,inner sep=0pt] at (2,0) {};
\end{tikzpicture}&=\begin{tikzpicture}[baseline=-0.5ex,scale=0.8]
        \draw[ultra thick,black] (-1,0) -- (0,0) -- (1,0);
        \node[below=4pt] at (-1,0) {$\ell^*_{ij}$};
         \node[below=4pt] at (0,0) {$y_a$};
         \node[below=4pt] at (1,0) {$y_b$};
          \node[rectangle, draw, fill=black, minimum width=5 pt, minimum height=5 pt,inner sep=0pt] (pt1) at (-1,0) {};
        \node[rectangle, draw, fill=black, minimum width=5 pt, minimum height=5 pt,inner sep=0pt] (A) at (0,0) {};
        \node[fill=black, circle, minimum size=5pt, inner sep=0pt, line width=1pt, draw=black] (B) at (1,0) {};
\end{tikzpicture}.
\label{eq:cb_1}
\end{align}
\item $ij$ and $kl$ are non-crossing chords
\begin{align}
 \begin{tikzpicture}[baseline=-0.5ex,scale=0.8]
        \draw[ultra thick,black] (0,0) -- (1,0);
        \draw[ultra thick,green] (1,0)--(2,0);
        \draw[ultra thick,black]  (-1,0)--(0,0);
        \node[below=4pt] at (2,0) {$\ell^*_{kl}$};
        \node[below=4pt] at (-1,0) {$\ell^*_{ij}$};
         \node[below=4pt] at (0,0) {$y_a$};
         \node[below=4pt] at (1,0) {$y_b$};
          \node[rectangle, draw, fill=black, minimum width=5 pt, minimum height=5 pt,inner sep=0pt] (pt1) at (-1,0) {};
        \node[rectangle, draw, fill=black, minimum width=5 pt, minimum height=5 pt,inner sep=0pt] (A) at (0,0) {};
        \node[fill=black, circle, minimum size=5pt, inner sep=0pt, line width=1pt, draw=black] (B) at (1,0) {};
        \node[rectangle, draw, fill=black, minimum width=5 pt, minimum height=5 pt,inner sep=0pt] at (2,0) {};
\end{tikzpicture} &=  \begin{tikzpicture}[baseline=-0.5ex,scale=0.8]
        \draw[ultra thick,black] (-1,0) -- (0,0) -- (1,0);
        \node[below=4pt] at (-1,0) {$\ell^*_{ij}$};
         \node[below=4pt] at (0,0) {$y_a$};
         \node[below=4pt] at (1,0) {$y_b$};
          \node[rectangle, draw, fill=black, minimum width=5 pt, minimum height=5 pt,inner sep=0pt] (pt1) at (-1,0) {};
        \node[rectangle, draw, fill=black, minimum width=5 pt, minimum height=5 pt,inner sep=0pt] (A) at (0,0) {};
        \node[fill=black, circle, minimum size=5pt, inner sep=0pt, line width=1pt, draw=black] (B) at (1,0) {};
\end{tikzpicture} , \notag \\
 \begin{tikzpicture}[baseline=-0.5ex,scale=0.8]
        \draw[ultra thick,black] (0,0) -- (1,0);
        \draw[ultra thick,red] (1,0)--(2,0);
        \draw[ultra thick,black]  (-1,0)--(0,0);
        \node[below=4pt] at (2,0) {$\ell^*_{kl}$};
        \node[below=4pt] at (-1,0) {$\ell^*_{ij}$};
         \node[below=4pt] at (0,0) {$y_a$};
         \node[below=4pt] at (1,0) {$y_b$};
          \node[rectangle, draw, fill=black, minimum width=5 pt, minimum height=5 pt,inner sep=0pt] (pt1) at (-1,0) {};
        \node[rectangle, draw, fill=black, minimum width=5 pt, minimum height=5 pt,inner sep=0pt] (A) at (0,0) {};
        \node[fill=black, circle, minimum size=5pt, inner sep=0pt, line width=1pt, draw=black] (B) at (1,0) {};
        \node[rectangle, draw, fill=black, minimum width=5 pt, minimum height=5 pt,inner sep=0pt] at (2,0) {};
\end{tikzpicture} &=0.
\label{eq:cb_2}
\end{align}
\item $ij$ and $jk$ are kissing chords such that in the cyclic order they appear as $jki$
\begin{align}
 \begin{tikzpicture}[baseline=-0.5ex,scale=0.8]
        \draw[ultra thick,black] (0,0) -- (1,0)--(2,0);
        \draw[ultra thick,green] (1,0)--(2,0);
        \draw[ultra thick,black]  (-1,0)--(0,0);
        \node[below=4pt] at (2,0) {$\ell^*_{jk}$};
        \node[below=4pt] at (-1,0) {$\ell^*_{ij}$};
         \node[below=4pt] at (0,0) {$y_a$};
         \node[below=4pt] at (1,0) {$y_b$};
          \node[rectangle, draw, fill=black, minimum width=5 pt, minimum height=5 pt,inner sep=0pt] (pt1) at (-1,0) {};
        \node[rectangle, draw, fill=black, minimum width=5 pt, minimum height=5 pt,inner sep=0pt] (A) at (0,0) {};
        \node[fill=black, circle, minimum size=5pt, inner sep=0pt, line width=1pt, draw=black] (B) at (1,0) {};
        \node[rectangle, draw, fill=black, minimum width=5 pt, minimum height=5 pt,inner sep=0pt] at (2,0) {};
\end{tikzpicture} &=\frac{4S_{x_ix_{i+1}x_j\ell^*_{jk}}(y_b-q^-_{ii+1j\ell^*_{jk}})^2(y_a-\ell^*_{ij})^2}{(y_b-x_i)^2(y_b-x_{i+1})^2(y_b-x_{j})^2(y_b-\ell^*_{jk})^2(y_b-y_a)^2},\notag \\
\begin{tikzpicture}[baseline=-0.5ex,scale=0.8]
        \draw[ultra thick,black] (0,0) -- (1,0)--(2,0);
        \draw[ultra thick,red] (1,0)--(2,0);
        \draw[ultra thick,black]  (-1,0)--(0,0);
        \node[below=4pt] at (2,0) {$\ell^*_{jk}$};
        \node[below=4pt] at (-1,0) {$\ell^*_{ij}$};
         \node[below=4pt] at (0,0) {$y_a$};
         \node[below=4pt] at (1,0) {$y_b$};
          \node[rectangle, draw, fill=black, minimum width=5 pt, minimum height=5 pt,inner sep=0pt] (pt1) at (-1,0) {};
        \node[rectangle, draw, fill=black, minimum width=5 pt, minimum height=5 pt,inner sep=0pt] (A) at (0,0) {};
        \node[fill=black, circle, minimum size=5pt, inner sep=0pt, line width=1pt, draw=black] (B) at (1,0) {};
        \node[rectangle, draw, fill=black, minimum width=5 pt, minimum height=5 pt,inner sep=0pt] at (2,0) {};
\end{tikzpicture}&=\frac{4S_{x_ix_{i+1}\ell^*_{jk}x_{j+1}}(y_b-q^+_{ii+1\ell^*_{jk}j+1})^2(y_a-\ell^*_{ij})^2}{(y_b-x_i)^2(y_b-x_{i+1})^2(y_b-\ell^*_{jk})^2(y_b-x_{j+1})^2(y_b-y_a)^2}.
\label{eq:cb_3}
\end{align}
\item $ij$ and $jk$ are kissing chords such that in the cyclic order they appear as $jik$
\begin{align}
 \begin{tikzpicture}[baseline=-0.5ex,scale=0.8]
        \draw[ultra thick,black] (0,0) -- (1,0)--(2,0);
        \draw[ultra thick,green] (1,0)--(2,0);
        \draw[ultra thick,black]  (-1,0)--(0,0);
        \node[below=4pt] at (2,0) {$\ell^*_{jk}$};
        \node[below=4pt] at (-1,0) {$\ell^*_{ij}$};
         \node[below=4pt] at (0,0) {$y_a$};
         \node[below=4pt] at (1,0) {$y_b$};
          \node[rectangle, draw, fill=black, minimum width=5 pt, minimum height=5 pt,inner sep=0pt] (pt1) at (-1,0) {};
        \node[rectangle, draw, fill=black, minimum width=5 pt, minimum height=5 pt,inner sep=0pt] (A) at (0,0) {};
        \node[fill=black, circle, minimum size=5pt, inner sep=0pt, line width=1pt, draw=black] (B) at (1,0) {};
        \node[rectangle, draw, fill=black, minimum width=5 pt, minimum height=5 pt,inner sep=0pt] at (2,0) {};
\end{tikzpicture} &=\frac{4S_{x_ix_{i+1}\ell^*_{jk}x_{j+1}}(y_b-q^-_{ii+1\ell^*_{jk}j+1})^2(y_a-\ell^*_{ij})^2}{(y_b-x_i)^2(y_b-x_{i+1})^2(y_b-\ell^*_{jk})^2(y_b-x_{j+1})^2(y_b-y_a)^2}, \notag \\
\begin{tikzpicture}[baseline=-0.5ex,scale=0.8]
        \draw[ultra thick,black] (0,0) -- (1,0)--(2,0);
        \draw[ultra thick,red] (1,0)--(2,0);
        \draw[ultra thick,black]  (-1,0)--(0,0);
        \node[below=4pt] at (2,0) {$\ell^*_{jk}$};
        \node[below=4pt] at (-1,0) {$\ell^*_{ij}$};
         \node[below=4pt] at (0,0) {$y_a$};
         \node[below=4pt] at (1,0) {$y_b$};
          \node[rectangle, draw, fill=black, minimum width=5 pt, minimum height=5 pt,inner sep=0pt] (pt1) at (-1,0) {};
        \node[rectangle, draw, fill=black, minimum width=5 pt, minimum height=5 pt,inner sep=0pt] (A) at (0,0) {};
        \node[fill=black, circle, minimum size=5pt, inner sep=0pt, line width=1pt, draw=black] (B) at (1,0) {};
        \node[rectangle, draw, fill=black, minimum width=5 pt, minimum height=5 pt,inner sep=0pt] at (2,0) {};
\end{tikzpicture}&=\frac{4S_{x_ix_{i+1}x_j\ell^*_{jk}}(y_b-q^+_{ii+1j\ell^*_{jk}})^2(y_a-\ell^*_{ij})^2}{(y_a-x_i)^2(y_b-x_{i+1})^2(y_b-x_{j})^2(y_b-\ell^*_{jk})^2(y_b-y_a)^2}.
\label{eq:cb_4}
\end{align}
\item $ij$ and $kl$ coincide
\begin{align}
 \begin{tikzpicture}[baseline=-0.5ex,scale=0.8]
        \draw[ultra thick,black] (0,0) -- (1,0)--(2,0);
        \draw[ultra thick,green] (1,0)--(2,0);
        \draw[ultra thick,black]  (-1,0)--(0,0);
        \node[below=4pt] at (2,0) {$\ell^*_{ij}$};
        \node[below=4pt] at (-1,0) {$\ell^*_{ij}$};
         \node[below=4pt] at (0,0) {$y_a$};
         \node[below=4pt] at (1,0) {$y_b$};
          \node[rectangle, draw, fill=black, minimum width=5 pt, minimum height=5 pt,inner sep=0pt] (pt1) at (-1,0) {};
        \node[rectangle, draw, fill=black, minimum width=5 pt, minimum height=5 pt,inner sep=0pt] (A) at (0,0) {};
        \node[fill=black, circle, minimum size=5pt, inner sep=0pt, line width=1pt, draw=black] (B) at (1,0) {};
        \node[rectangle, draw, fill=black, minimum width=5 pt, minimum height=5 pt,inner sep=0pt] at (2,0) {};
\end{tikzpicture} &= \frac{4(x_{i}-x_{j+1})^2(y_a-\ell^*_{ij})^2}{(y_b-x_{i})^2(y_b-\ell^*_{ij})^2(y_b-x_{j+1})^2(y_b-y_a)^2}   \notag \\
& +\frac{4(x_{i+1}-x_j)^2(y_a-\ell^*_{ij})^2}{(y_b-x_{i+1})^2(y_b-\ell^*_{ij})^2(y_b-x_j)^2(y_b-y_a)^2},  \notag \\
\begin{tikzpicture}[baseline=-0.5ex,scale=0.8]
        \draw[ultra thick,black] (0,0) -- (1,0)--(2,0);
        \draw[ultra thick,red] (1,0)--(2,0);
        \draw[ultra thick,black]  (-1,0)--(0,0);
        \node[below=4pt] at (2,0) {$\ell^*_{ij}$};
        \node[below=4pt] at (-1,0) {$\ell^*_{ij}$};
         \node[below=4pt] at (0,0) {$y_a$};
         \node[below=4pt] at (1,0) {$y_b$};
          \node[rectangle, draw, fill=black, minimum width=5 pt, minimum height=5 pt,inner sep=0pt] (pt1) at (-1,0) {};
        \node[rectangle, draw, fill=black, minimum width=5 pt, minimum height=5 pt,inner sep=0pt] (A) at (0,0) {};
        \node[fill=black, circle, minimum size=5pt, inner sep=0pt, line width=1pt, draw=black] (B) at (1,0) {};
        \node[rectangle, draw, fill=black, minimum width=5 pt, minimum height=5 pt,inner sep=0pt] at (2,0) {};
\end{tikzpicture}&=-\frac{4(x_{i}-x_{j})^2(y_a-\ell^*_{ij})^2}{(y_b-x_{i})^2(y_b-\ell^*_{ij})^2(y_b-x_{j})^2(y_b-y_a)^2} \notag \\
&-\frac{4(x_{i+1}-x_{j+1})^2(y_a-\ell^*_{ij})^2}{(y_b-x_{i+1})^2(y_b-\ell^*_{ij})^2(y_b-x_{j+1})^2(y_b-y_a)^2}.
\label{eq:cb_5}\end{align}
\end{itemize}
Inserting these results into \eqref{eq:box_coeffs} and reorganising the sums we find the following recursion for the coefficients of the chiral box
\begin{align}
 \begin{tikzpicture}[baseline=-0.5ex,scale=0.8]
        \draw[ultra thick,red] (0,0) -- (1,0)--(1.2,0);
        \draw[ultra thick,red] (1.8,0) -- (2,0) -- (3,0);
        \node[fill=black, circle, minimum size=4pt, inner sep=0pt, line width=1pt, draw=black] at (0,0) {};
        \node[fill=black, circle, minimum size=4pt, inner sep=0pt, line width=1pt, draw=black] at (1,0) {};
        \node[fill=black, circle, minimum size=4pt, inner sep=0pt, line width=1pt, draw=black] at (2,0) {};
        \node[rectangle, draw, fill=black, minimum width=5 pt, minimum height=5 pt,inner sep=0pt] at (3,0) {};
        \node[] at (1.55,0) {\textcolor{red}{{\small \ldots}}};
         \node[below=4pt] at (0,0) {$y_1$};
         \node[below=4pt] at (1,0) {$y_2$};
         \node[below=4pt] at (2,0) {$y_{L}$};
         \node[below=4pt] at (3,0) {$\ell^*_{ij}$};
\end{tikzpicture}=\sum_{\ell^*_{kl} \in \mathcal{V}^-} \begin{tikzpicture}[baseline=-0.5ex,scale=0.8]
        \draw[ultra thick,red] (0,0) -- (1,0)--(1.2,0);
        \draw[ultra thick,red] (1.8,0) -- (2,0) -- (3,0);
        \node[fill=black, circle, minimum size=4pt, inner sep=0pt, line width=1pt, draw=black] at (0,0) {};
        \node[fill=black, circle, minimum size=4pt, inner sep=0pt, line width=1pt, draw=black] at (1,0) {};
        \node[fill=black, circle, minimum size=4pt, inner sep=0pt, line width=1pt, draw=black] at (2,0) {};
        \node[rectangle, draw, fill=black, minimum width=5 pt, minimum height=5 pt,inner sep=0pt] at (3,0) {};
        \node[] at (1.55,0) {\textcolor{red}{{\small \ldots}}};
         \node[below=4pt] at (0,0) {$y_1$};
         \node[below=4pt] at (1,0) {$y_2$};
         \node[below=4pt] at (2,0) {$y_{L-1}$};
         \node[below=4pt] at (3,0) {$\ell^*_{kl}$};
\end{tikzpicture} \wedge \begin{tikzpicture}[baseline=-0.5ex,scale=0.8]
        \draw[ultra thick,black] (-1,0) -- (1,0);
        \draw[ultra thick,red] (1,0)--(2,0);
        \node[below=4pt] at (2,0) {$\ell^*_{ij}$};
        \node[below=4pt] at (-1,0) {$\ell^*_{kl}$};
         \node[below=4pt] at (0,0) {$y_{L-1}$};
         \node[below=4pt] at (1,0) {$y_{L}$};
          \node[rectangle, draw, fill=black, minimum width=5 pt, minimum height=5 pt,inner sep=0pt] (pt1) at (-1,0) {};
        \node[rectangle, draw, fill=black, minimum width=5 pt, minimum height=5 pt,inner sep=0pt] (A) at (0,0) {};
        \node[fill=black, circle, minimum size=5pt, inner sep=0pt, line width=1pt, draw=black] (B) at (1,0) {};
        \node[rectangle, draw, fill=black, minimum width=5 pt, minimum height=5 pt,inner sep=0pt] at (2,0) {};
\end{tikzpicture}.
\label{eq:recursion}
\end{align}
It is useful to define the following factor for $a<b$
\begin{align}
 \begin{tikzpicture}[baseline=-0.5ex,scale=0.75]
        \draw[ultra thick,black] (-1,0) -- (1,0);
        \draw[ultra thick,red] (1,0)--(1.2,0);
        \draw[ultra thick,red] (1.8,0) -- (2,0) -- (3,0);
        \node[rectangle, draw, fill=black, minimum width=5 pt, minimum height=5 pt,inner sep=0pt] at (-1,0) {};
         \node[rectangle, draw, fill=black, minimum width=5 pt, minimum height=5 pt,inner sep=0pt] at (0,0) {};
        \node[fill=black, circle, minimum size=4pt, inner sep=0pt, line width=1pt, draw=black] at (1,0) {};
        \node[fill=black, circle, minimum size=4pt, inner sep=0pt, line width=1pt, draw=black] at (2,0) {};
        \node[rectangle, draw, fill=black, minimum width=5 pt, minimum height=5 pt,inner sep=0pt] at (3,0) {};
        \node[] at (1.55,0) {\textcolor{red}{{\small \ldots}}};
         \node[below=4pt] at (-1,0) {$\ell^*_{kl}$};
         \node[below=4pt] at (0,0) {$y_a$};
         \node[below=4pt] at (1,0) {$y_{a+1}$};
         \node[below=4pt] at (2,0) {$y_{b}$};
         \node[below=4pt] at (3,0) {$\ell^*_{ij}$};
\end{tikzpicture} =  \sum_{\ell^*_{pq} \in \mathcal{V}^-} \ldots \sum_{\ell^*_{mn} \in \mathcal{V}^-} \begin{tikzpicture}[baseline=-0.5ex,scale=0.75]
        \draw[ultra thick,black] (-1,0) -- (1,0);
        \draw[ultra thick,red] (1,0)--(2,0);
        \node[below=4pt] at (2,0) {$\ell^*_{pq}$};
        \node[below=4pt] at (-1,0) {$\ell^*_{kl}$};
         \node[below=4pt] at (0,0) {$y_a$};
         \node[below=4pt] at (1,0) {$y_{a+1}$};
          \node[rectangle, draw, fill=black, minimum width=5 pt, minimum height=5 pt,inner sep=0pt] (pt1) at (-1,0) {};
        \node[rectangle, draw, fill=black, minimum width=5 pt, minimum height=5 pt,inner sep=0pt] (A) at (0,0) {};
        \node[fill=black, circle, minimum size=5pt, inner sep=0pt, line width=1pt, draw=black] (B) at (1,0) {};
        \node[rectangle, draw, fill=black, minimum width=5 pt, minimum height=5 pt,inner sep=0pt] at (2,0) {};
\end{tikzpicture} \wedge \ldots \wedge \begin{tikzpicture}[baseline=-0.5ex,scale=0.75]
        \draw[ultra thick,black] (-1,0) -- (1,0);
        \draw[ultra thick,red] (1,0)--(2,0);
        \node[below=4pt] at (2,0) {$\ell^*_{ij}$};
        \node[below=4pt] at (-1,0) {$\ell^*_{mn}$};
         \node[below=4pt] at (0,0) {$y_{b-1}$};
         \node[below=4pt] at (1,0) {$y_{b}$};
          \node[rectangle, draw, fill=black, minimum width=5 pt, minimum height=5 pt,inner sep=0pt] (pt1) at (-1,0) {};
        \node[rectangle, draw, fill=black, minimum width=5 pt, minimum height=5 pt,inner sep=0pt] (A) at (0,0) {};
        \node[fill=black, circle, minimum size=5pt, inner sep=0pt, line width=1pt, draw=black] (B) at (1,0) {};
        \node[rectangle, draw, fill=black, minimum width=5 pt, minimum height=5 pt,inner sep=0pt] at (2,0) {};
\end{tikzpicture},
\label{eq:middle_factor}
\end{align}
such that inserting the recursion \eqref{eq:recursion} into the formula \eqref{eq:ladder_int} for the $(L+1)$-loop negative ladder we arrive at our main result
\begin{align}
 \begin{tikzpicture}[baseline=-0.5ex,scale=0.7]
        \draw[ultra thick,red] (0,0) -- (1,0)--(1.2,0);
        \draw[ultra thick,red] (1.8,0) -- (2,0) -- (3,0);
        \node[fill=black, circle, minimum size=4pt, inner sep=0pt, line width=1pt, draw=black] (A) at (0,0) {};
        \node[fill=black, circle, minimum size=4pt, inner sep=0pt, line width=1pt, draw=black] (B) at (1,0) {};
        \node[fill=black, circle, minimum size=4pt, inner sep=0pt, line width=1pt, draw=black] (C) at (2,0) {};
        \node[fill=black, circle, minimum size=4pt, inner sep=0pt, line width=1pt, draw=black] (D) at (3,0) {};
        \node[] at (1.55,0) {\textcolor{red}{{\small \ldots}}};
         \node[below=4pt] at (0,0) {$y_1$};
         \node[below=4pt] at (1,0) {$y_2$};
         \node[below=4pt] at (3,0) {$y_{L+1}$};
\end{tikzpicture}=\sum_{\ell^*_{kl} \in \mathcal{V}^- } \sum_{\ell^*_{ij} \in \mathcal{V}^- }\begin{tikzpicture}[baseline=-0.5ex,scale=0.6]
        \draw[ultra thick,red] (0,0) -- (1,0);
         \node[below=4pt] at (0,0) {$y_1$};
         \node[below=4pt] at (1,0) {$\ell^*_{kl}$};
        \node[fill=black, circle, minimum size=5pt, inner sep=0pt, line width=1pt, draw=black] (A) at (0,0) {};
        \node[rectangle, draw, fill=black, minimum width=5 pt, minimum height=5 pt,inner sep=0pt] (B) at (1,0) {};
\end{tikzpicture} \wedge  \begin{tikzpicture}[baseline=-0.5ex,scale=0.7]
        \draw[ultra thick,black] (-1,0) -- (1,0);
        \draw[ultra thick,red] (1,0)--(1.2,0);
        \draw[ultra thick,red] (1.8,0) -- (2,0) -- (3,0);
        \node[rectangle, draw, fill=black, minimum width=5 pt, minimum height=5 pt,inner sep=0pt] at (-1,0) {};
         \node[rectangle, draw, fill=black, minimum width=5 pt, minimum height=5 pt,inner sep=0pt] at (0,0) {};
        \node[fill=black, circle, minimum size=4pt, inner sep=0pt, line width=1pt, draw=black] at (1,0) {};
        \node[fill=black, circle, minimum size=4pt, inner sep=0pt, line width=1pt, draw=black] at (2,0) {};
        \node[rectangle, draw, fill=black, minimum width=5 pt, minimum height=5 pt,inner sep=0pt] at (3,0) {};
        \node[] at (1.55,0) {\textcolor{red}{{\small \ldots}}};
         \node[below=4pt] at (-1,0) {$\ell^*_{kl}$};
         \node[below=4pt] at (0,0) {$y_1$};
         \node[below=4pt] at (1,0) {$y_{2}$};
         \node[below=4pt] at (2,0) {$y_{L}$};
         \node[below=4pt] at (3,0) {$\ell^*_{ij}$};
\end{tikzpicture}   \wedge \begin{tikzpicture}[baseline=-0.5ex,scale=0.7]
        \draw[ultra thick,black] (-1,0) -- (1,0);
        \node[below=4pt] at (-1,0) {$\ell^*_{ij}$};
         \node[below=4pt] at (0,0) {$y_{L}$};
         \node[below=4pt] at (1,0) {$y_{L+1}$};
          \node[rectangle, draw, fill=black, minimum width=5 pt, minimum height=5 pt,inner sep=0pt] (pt1) at (-1,0) {};
        \node[rectangle, draw, fill=black, minimum width=5 pt, minimum height=5 pt,inner sep=0pt] (A) at (0,0) {};
        \node[fill=black, circle, minimum size=5pt, inner sep=0pt, line width=1pt, draw=black] (B) at (1,0) {};
\end{tikzpicture},
\label{eq:main_result}
\end{align}
where all terms appearing on the right hand side have been defined in \eqref{eq:subamps_neg}, \eqref{eq:chiral_box}, \eqref{eq:cb_1}-\eqref{eq:cb_5} and \eqref{eq:middle_factor}. The formula for an arbitrary ladder can be obtained by simultaneously flipping red edges to green edges, and the corresponding sum range from $\mathcal{V}^-$ to $\mathcal{V}^+$ in the above formulae. We have checked the formula \eqref{eq:main_result} is symmetric under exchanging $y_{a} \leftrightarrow y_{y_{L-a+2}}$ which serves as a non-trivial consistency check. 

It is useful to see how this formula behaves in the simplest cases beginning at four-points. Recall, the four-point case is drastically simplified due to the one-loop region being covered by a single sign pattern (i.e. one-loop chamber) such that we have the graphical identities
\begin{align}
\begin{tikzpicture}[baseline=-0.5ex,scale=0.8]
        \draw[ultra thick,red]  (-1,0)--(-2,0);
        \node[below=4pt] at (-1,0) {{\small $\ell^*_{ij}$}};
        \node[below=4pt] at (-2,0) {{\small $y_1$}};
         \node[rectangle, draw, fill=black, minimum width=5 pt, minimum height=5 pt,inner sep=0pt] (pt1) at (-1,0) {};
        \node[fill=black, circle, minimum size=5pt, inner sep=0pt, line width=1pt, draw=black] at (-2,0) {};
\end{tikzpicture} & =\begin{tikzpicture}[baseline=-0.5ex,scale=0.8]
         \node[below=4pt] at (0,0) {$y_1$};
        \node[fill=black, circle, minimum size=5pt, inner sep=0pt, line width=1pt, draw=black] (A) at (0,0) {};
\end{tikzpicture} , \quad \quad  \begin{tikzpicture}[baseline=-0.5ex,scale=0.8]
        \draw[ultra thick,black] (0,0) -- (1,0);
        \draw[ultra thick,red] (1,0)--(2,0);
        \draw[ultra thick,black]  (-1,0)--(0,0);
        \node[below=4pt] at (2,0) {$\ell^*_{kl}$};
        \node[below=4pt] at (-1,0) {$\ell^*_{ij}$};
         \node[below=4pt] at (0,0) {$y_b$};
         \node[below=4pt] at (1,0) {$y_a$};
          \node[rectangle, draw, fill=black, minimum width=5 pt, minimum height=5 pt,inner sep=0pt] (pt1) at (-1,0) {};
        \node[rectangle, draw, fill=black, minimum width=5 pt, minimum height=5 pt,inner sep=0pt] (A) at (0,0) {};
        \node[fill=black, circle, minimum size=5pt, inner sep=0pt, line width=1pt, draw=black] (B) at (1,0) {};
        \node[rectangle, draw, fill=black, minimum width=5 pt, minimum height=5 pt,inner sep=0pt] at (2,0) {};
\end{tikzpicture}=\begin{tikzpicture}[baseline=-0.5ex,scale=0.8]
        \draw[ultra thick,black] (-1,0) -- (0,0) -- (1,0);
        \node[below=4pt] at (-1,0) {$\ell^*_{ij}$};
         \node[below=4pt] at (0,0) {$y_b$};
         \node[below=4pt] at (1,0) {$y_a$};
          \node[rectangle, draw, fill=black, minimum width=5 pt, minimum height=5 pt,inner sep=0pt] (pt1) at (-1,0) {};
        \node[rectangle, draw, fill=black, minimum width=5 pt, minimum height=5 pt,inner sep=0pt] (A) at (0,0) {};
        \node[fill=black, circle, minimum size=5pt, inner sep=0pt, line width=1pt, draw=black] (B) at (1,0) {};
\end{tikzpicture}.
\end{align}
This means that for example the four-loop negative ladder can be written as 
\begin{align}
\begin{tikzpicture}[baseline=-0.5ex,scale=0.6]
         \node[below=4pt] at (0,0) {$y_1$};
         \node[below=4pt] at (1,0) {$y_2$};
         \node[below=4pt] at (2,0) {$y_3$};
          \node[below=4pt] at (3,0) {$y_4$};
         \draw[ultra thick,red]  (0,0)--(1,0);
         \draw[ultra thick,red]  (1,0)--(2,0);
         \draw[ultra thick,red]  (2,0)--(3,0);
        \node[fill=black, circle, minimum size=5pt, inner sep=0pt, line width=1pt, draw=black] at (0,0) {};
        \node[fill=black, circle, minimum size=5pt, inner sep=0pt, line width=1pt, draw=black] at (1,0) {};
        \node[fill=black, circle, minimum size=5pt, inner sep=0pt, line width=1pt, draw=black] at (2,0) {};
        \node[fill=black, circle, minimum size=5pt, inner sep=0pt, line width=1pt, draw=black] at (3,0) {};
\end{tikzpicture} & = \begin{tikzpicture}[baseline=-0.5ex,scale=0.8]
        \node[below=4pt] at (-2,0) {{\small $y_1$}};
        \node[fill=black, circle, minimum size=5pt, inner sep=0pt, line width=1pt, draw=black] at (-2,0) {};
\end{tikzpicture} \wedge  \sum_{\ell^*_{ij} \in \vertsRed{}} \begin{tikzpicture}[baseline=-0.5ex,scale=0.6]
        \draw[ultra thick,black]  (-3,0)--(-2,0);
        \draw[ultra thick,black]  (-1,0)--(-2,0);
        \node[below=4pt] at (-3,0) {{\small $\ell^*_{ij}$}};
        \node[below=4pt] at (-2,0) {{\small $y_1$}};
        \node[below=4pt] at (-1,0) {{\small $y_2$}};
        \node[rectangle, draw, fill=black, minimum width=5 pt, minimum height=5 pt,inner sep=0pt] (pt1) at (-3,0) {};
        \node[rectangle, draw, fill=black, minimum width=5 pt, minimum height=5 pt,inner sep=0pt] (pt2) at (-2,0) {};
        \node[fill=black, circle, minimum size=5pt, inner sep=0pt, line width=1pt, draw=black] at (-1,0) {};
\end{tikzpicture} \wedge \sum_{\ell^*_{kl} \in \vertsRed{}} \begin{tikzpicture}[baseline=-0.5ex,scale=0.6]
        \draw[ultra thick,black]  (-3,0)--(-2,0);
        \draw[ultra thick,black]  (-1,0)--(-2,0);
        \node[below=4pt] at (-3,0) {{\small $\ell^*_{kl}$}};
        \node[below=4pt] at (-2,0) {{\small $y_2$}};
        \node[below=4pt] at (-1,0) {{\small $y_3$}};
        \node[rectangle, draw, fill=black, minimum width=5 pt, minimum height=5 pt,inner sep=0pt] (pt1) at (-3,0) {};
        \node[rectangle, draw, fill=black, minimum width=5 pt, minimum height=5 pt,inner sep=0pt] (pt2) at (-2,0) {};
        \node[fill=black, circle, minimum size=5pt, inner sep=0pt, line width=1pt, draw=black] at (-1,0) {};
\end{tikzpicture}\wedge \sum_{\ell^*_{mn} \in \vertsRed{}} \begin{tikzpicture}[baseline=-0.5ex,scale=0.6]
        \draw[ultra thick,black]  (-3,0)--(-2,0);
        \draw[ultra thick,black]  (-1,0)--(-2,0);
        \node[below=4pt] at (-3,0) {{\small $\ell^*_{mn}$}};
        \node[below=4pt] at (-2,0) {{\small $y_3$}};
        \node[below=4pt] at (-1,0) {{\small $y_4$}};
        \node[rectangle, draw, fill=black, minimum width=5 pt, minimum height=5 pt,inner sep=0pt] (pt1) at (-3,0) {};
        \node[rectangle, draw, fill=black, minimum width=5 pt, minimum height=5 pt,inner sep=0pt] (pt2) at (-2,0) {};
        \node[fill=black, circle, minimum size=5pt, inner sep=0pt, line width=1pt, draw=black] at (-1,0) {};
\end{tikzpicture}.
\end{align}
This matches the results for the four-point ladders originally found in \cite{Arkani-Hamed:2021iya}. Interestingly, in this case, the above formula can be immediately generalised to arbitrary trees in loop space, and any ladder can be written solely using chiral pentagons. 

\subsection{Alternative expansion}
There exists an alternative expansion of the ladder where instead of performing the recursion \eqref{eq:recursion} at one end of the ladder we instead perform the recursion at {\it both} ends until reaching some loop momentum $y_a$ in the middle. The solution to this recursion provides the following alternative form for the negative ladder
\begin{align}
 \begin{tikzpicture}[baseline=-0.5ex,scale=0.7]
        \draw[ultra thick,red] (0,0) -- (1,0)--(1.2,0);
        \draw[ultra thick,red] (1.8,0) -- (2,0) -- (3,0);
        \node[fill=black, circle, minimum size=4pt, inner sep=0pt, line width=1pt, draw=black] (A) at (0,0) {};
        \node[fill=black, circle, minimum size=4pt, inner sep=0pt, line width=1pt, draw=black] (B) at (1,0) {};
        \node[fill=black, circle, minimum size=4pt, inner sep=0pt, line width=1pt, draw=black] (C) at (2,0) {};
        \node[fill=black, circle, minimum size=4pt, inner sep=0pt, line width=1pt, draw=black] (D) at (3,0) {};
        \node[] at (1.55,0) {\textcolor{red}{{\small \ldots}}};
         \node[below=4pt] at (0,0) {$y_1$};
         \node[below=4pt] at (1,0) {$y_2$};
         \node[below=4pt] at (3,0) {$y_{L+1}$};
\end{tikzpicture}=&\sum_{\ell^*_{pq} \in \mathcal{V}^- } \sum_{\ell^*_{mn} \in \mathcal{V}^- }\begin{tikzpicture}[baseline=-0.5ex,scale=0.5]
        \draw[ultra thick,black] (0,0) -- (2,0);
         \node[below=4pt] at (0,0) {$y_1$};
         \node[below=4pt] at (1,0) {$y_2$};
         \node[below=4pt] at (2,0) {$\ell^*_{pq}$};
        \node[fill=black, circle, minimum size=5pt, inner sep=0pt, line width=1pt, draw=black] (A) at (0,0) {};
        \node[rectangle, draw, fill=black, minimum width=5 pt, minimum height=5 pt,inner sep=0pt] (B) at (1,0) {};
        \node[rectangle, draw, fill=black, minimum width=5 pt, minimum height=5 pt,inner sep=0pt] (B) at (2,0) {};
\end{tikzpicture} \wedge \begin{tikzpicture}[baseline=-0.5ex,scale=0.7]
        \draw[ultra thick,black] (3,0) -- (1,0);
        \draw[ultra thick,red] (1,0)--(0.8,0);
        \draw[ultra thick,red] (0.2,0) -- (0,0) -- (-1,0);
        \node[rectangle, draw, fill=black, minimum width=5 pt, minimum height=5 pt,inner sep=0pt] at (3,0) {};
         \node[rectangle, draw, fill=black, minimum width=5 pt, minimum height=5 pt,inner sep=0pt] at (2,0) {};
        \node[fill=black, circle, minimum size=4pt, inner sep=0pt, line width=1pt, draw=black] at (1,0) {};
        \node[fill=black, circle, minimum size=4pt, inner sep=0pt, line width=1pt, draw=black] at (0,0) {};
        \node[rectangle, draw, fill=black, minimum width=5 pt, minimum height=5 pt,inner sep=0pt] at (-1,0) {};
        \node[] at (0.55,0) {\textcolor{red}{{\small \ldots}}};
         \node[below=4pt] at (-1,0) {$\ell^*_{pq}$};
         \node[below=4pt] at (0,0) {$y_2$};
         \node[below=4pt] at (1,0) {$y_{a-1}$};
         \node[below=4pt] at (2,0) {$y_a$};
         \node[below=4pt] at (3,0) {$\ell^*_{mn}$};
\end{tikzpicture}  \wedge \begin{tikzpicture}[baseline=-0.5ex,scale=0.6]
        \draw[ultra thick,red] (-1,0) -- (1,0);
        \node[below=4pt] at (-1,0) {$\ell^*_{mn}$};
         \node[below=4pt] at (0,0) {$y_a$};
         \node[below=4pt] at (1,0) {$\ell^*_{kl}$};
          \node[rectangle, draw, fill=black, minimum width=5 pt, minimum height=5 pt,inner sep=0pt] (pt1) at (-1,0) {};
         \node[fill=black, circle, minimum size=5pt, inner sep=0pt, line width=1pt, draw=black] at (0,0) {};
         \node[rectangle, draw, fill=black, minimum width=5 pt, minimum height=5 pt,inner sep=0pt] at (1,0) {};
\end{tikzpicture} \wedge  \notag \\
&\sum_{\ell^*_{kl} \in \mathcal{V}^- } \sum_{\ell^*_{ij} \in \mathcal{V}^- }
\begin{tikzpicture}[baseline=-0.5ex,scale=0.7]
        \draw[ultra thick,black] (-1,0) -- (1,0);
        \draw[ultra thick,red] (1,0)--(1.2,0);
        \draw[ultra thick,red] (1.8,0) -- (2,0) -- (3,0);
        \node[rectangle, draw, fill=black, minimum width=5 pt, minimum height=5 pt,inner sep=0pt] at (-1,0) {};
         \node[rectangle, draw, fill=black, minimum width=5 pt, minimum height=5 pt,inner sep=0pt] at (0,0) {};
        \node[fill=black, circle, minimum size=4pt, inner sep=0pt, line width=1pt, draw=black] at (1,0) {};
        \node[fill=black, circle, minimum size=4pt, inner sep=0pt, line width=1pt, draw=black] at (2,0) {};
        \node[rectangle, draw, fill=black, minimum width=5 pt, minimum height=5 pt,inner sep=0pt] at (3,0) {};
        \node[] at (1.55,0) {\textcolor{red}{{\small \ldots}}};
         \node[below=4pt] at (-1,0) {$\ell^*_{kl}$};
         \node[below=4pt] at (0,0) {$y_a$};
         \node[below=4pt] at (1,0) {$y_{a+1}$};
         \node[below=4pt] at (2,0) {$y_{L}$};
         \node[below=4pt] at (3,0) {$\ell^*_{ij}$};
\end{tikzpicture}  \wedge \begin{tikzpicture}[baseline=-0.5ex,scale=0.7]
        \draw[ultra thick,black] (-1,0) -- (1,0);
        \node[below=4pt] at (-1,0) {$\ell^*_{ij}$};
         \node[below=4pt] at (0,0) {$y_{L}$};
         \node[below=4pt] at (1,0) {$y_{L+1}$};
          \node[rectangle, draw, fill=black, minimum width=5 pt, minimum height=5 pt,inner sep=0pt] (pt1) at (-1,0) {};
        \node[rectangle, draw, fill=black, minimum width=5 pt, minimum height=5 pt,inner sep=0pt] (A) at (0,0) {};
        \node[fill=black, circle, minimum size=5pt, inner sep=0pt, line width=1pt, draw=black] (B) at (1,0) {};
\end{tikzpicture},
\label{eq:alt_result}
\end{align}
where we have defined the factors for $a>b$ similar to \eqref{eq:middle_factor} as
 \begin{align}
 \begin{tikzpicture}[baseline=-0.5ex,scale=0.7]
        \draw[ultra thick,black] (3,0) -- (1,0);
        \draw[ultra thick,red] (1,0)--(0.8,0);
        \draw[ultra thick,red] (0.2,0) -- (0,0) -- (-1,0);
        \node[rectangle, draw, fill=black, minimum width=5 pt, minimum height=5 pt,inner sep=0pt] at (3,0) {};
         \node[rectangle, draw, fill=black, minimum width=5 pt, minimum height=5 pt,inner sep=0pt] at (2,0) {};
        \node[fill=black, circle, minimum size=4pt, inner sep=0pt, line width=1pt, draw=black] at (1,0) {};
        \node[fill=black, circle, minimum size=4pt, inner sep=0pt, line width=1pt, draw=black] at (0,0) {};
        \node[rectangle, draw, fill=black, minimum width=5 pt, minimum height=5 pt,inner sep=0pt] at (-1,0) {};
        \node[] at (0.55,0) {\textcolor{red}{{\small \ldots}}};
         \node[below=4pt] at (-1,0) {$\ell^*_{ij}$};
         \node[below=4pt] at (0,0) {$y_b$};
         \node[below=4pt] at (1,0) {$y_{a-1}$};
         \node[below=4pt] at (2,0) {$y_{a}$};
         \node[below=4pt] at (3,0) {$\ell^*_{kl}$};
\end{tikzpicture}=  \sum_{\ell^*_{pq} \in \mathcal{V}^-} \ldots \sum_{\ell^*_{mn} \in \mathcal{V}^-} \begin{tikzpicture}[baseline=-0.5ex,scale=0.75]
        \draw[ultra thick,black] (2,0) -- (0,0);
        \draw[ultra thick,red] (0,0)--(-1,0);
        \node[below=4pt] at (2,0) {$\ell^*_{mn}$};
        \node[below=4pt] at (1,0) {$y_{b+1}$};
        \node[below=4pt] at (0,0) {$y_b$};
        \node[below=4pt] at (-1,0) {$\ell^*_{ij}$};
          \node[rectangle, draw, fill=black, minimum width=5 pt, minimum height=5 pt,inner sep=0pt] (pt1) at (2,0) {};
        \node[rectangle, draw, fill=black, minimum width=5 pt, minimum height=5 pt,inner sep=0pt] (A) at (1,0) {};
        \node[fill=black, circle, minimum size=5pt, inner sep=0pt, line width=1pt, draw=black] (B) at (0,0) {};
        \node[rectangle, draw, fill=black, minimum width=5 pt, minimum height=5 pt,inner sep=0pt] at (-1,0) {};
\end{tikzpicture} \wedge \ldots \wedge \begin{tikzpicture}[baseline=-0.5ex,scale=0.75]
        \draw[ultra thick,black] (2,0) -- (0,0);
        \draw[ultra thick,red] (0,0)--(-1,0);
        \node[below=4pt] at (2,0) {$\ell^*_{kl}$};
        \node[below=4pt] at (1,0) {$y_{a}$};
        \node[below=4pt] at (0,0) {$y_{a-1}$};
        \node[below=4pt] at (-1,0) {$\ell^*_{pq}$};
          \node[rectangle, draw, fill=black, minimum width=5 pt, minimum height=5 pt,inner sep=0pt] (pt1) at (2,0) {};
        \node[rectangle, draw, fill=black, minimum width=5 pt, minimum height=5 pt,inner sep=0pt] (A) at (1,0) {};
        \node[fill=black, circle, minimum size=5pt, inner sep=0pt, line width=1pt, draw=black] (B) at (0,0) {};
        \node[rectangle, draw, fill=black, minimum width=5 pt, minimum height=5 pt,inner sep=0pt] at (-1,0) {};
\end{tikzpicture},
\end{align}
and the canonical forms of the new factors appearing on the vertex $y_a$ are given explicitly by
\begin{align}
 \begin{tikzpicture}[baseline=-0.5ex,scale=0.7]
        \draw[ultra thick,red] (-1,0) -- (1,0);
         \node[below=4pt] at (0,0) {$y_a$};
         \node[below=4pt] at (-1,0) {$\ell^*_{kl}$};
         \node[below=4pt] at (1,0) {$\ell^*_{ij}$};
        \node[fill=black, circle, minimum size=5pt, inner sep=0pt, line width=1pt, draw=black] (A) at (0,0) {};
        \node[rectangle, draw, fill=black, minimum width=5 pt, minimum height=5 pt,inner sep=0pt] (B) at (1,0) {};
        \node[rectangle, draw, fill=black, minimum width=5 pt, minimum height=5 pt,inner sep=0pt] (B) at (-1,0) {};
\end{tikzpicture}&= \sum_{\ell^*_{mn} \text{{\tiny cross }} \ell^*_{kl}}  \begin{tikzpicture}[baseline=-0.5ex,scale=0.7]
        \draw[ultra thick,black] (-1,0) -- (1,0);
        \draw[ultra thick,red] (1,0)--(2,0);
        \node[below=4pt] at (2,0) {$\ell^*_{ij}$};
        \node[below=4pt] at (-1,0) {$\ell^*_{mn}$};
         \node[below=4pt] at (0,0) {$\ell^*_{kl}$};
         \node[below=4pt] at (1,0) {$y_a$};
          \node[rectangle, draw, fill=black, minimum width=5 pt, minimum height=5 pt,inner sep=0pt] (pt1) at (-1,0) {};
        \node[rectangle, draw, fill=black, minimum width=5 pt, minimum height=5 pt,inner sep=0pt] (A) at (0,0) {};
        \node[fill=black, circle, minimum size=5pt, inner sep=0pt, line width=1pt, draw=black] (B) at (1,0) {};
        \node[rectangle, draw, fill=black, minimum width=5 pt, minimum height=5 pt,inner sep=0pt] at (2,0) {};
\end{tikzpicture},
\end{align}
where the sum is over all chords $mn$ of the $n$-gon which cross $kl$. The expansion of the form \eqref{eq:alt_result} is a crucial ingredient required to find integrated finite quantities, as explained in \cite{Chicherin:2024hes}.

\section{Conclusion and outlook}
\label{sec:conc}
In this paper we have studied the geometries associated to ladders in loop space defined by relaxing mutual positivity constraints between loop momenta in the definition of the momentum amplituhedron for MHV$_n$ scattering amplitudes in planar $\mathcal{N}=4$ SYM. Our main result \eqref{eq:main_result} is a formula for an arbitrary positive/negative ladder in loop space for any multiplicity $n$. Remarkably, just as was the case for the two-loop positive ladder, already studied in \cite{Ferro:2024vwn}, we find that the canonical form can be written in a term-wise factorised form extending the idea of fibrations of fibrations to all loop orders for ladder type geometries.

There is a number of interesting future research directions to pursue. The ladder geometries studied in this paper are the simplest family of graphs in loop space and it would be interesting to see how our results extend to more general topologies. In particular, the next simplest case to consider would be arbitrary trees in loop space. At four-points the extension of the ladder results to general trees is known \cite{Arkani-Hamed:2021iya}, however, we expect additional complications for $n>4$. As opposed to the case of ladders where each loop variable is connected to at most two other loops, for general tree graphs one needs to consider vertices with higher valency. We believe that the fibration of fibration idea will provide a good starting point for general trees, however, in order to find contributions from higher valency vertices in loop space, the canonical forms for all one-loop chambers, currently unknown, will be required.  

In this paper we have considered the negative ladders only at the level of the integrand, however, recent progress has been made towards integrating these objects at four and five-points \cite{Arkani-Hamed:2021iya,Brown:2023mqi,Chicherin:2024hes}. It is therefore a natural next step to integrate the expressions that we wrote in formula \eqref{eq:main_result} over $L-1$ loop points, that compute an $n$-point Wilson loop with one Lagrangian insertion. As for the case of four and five points, this quantity is finite, but for $n>5$ it will start depending on non-trivial cross-ratios of external kinematics, leading to a plethora of new interesting structures. Integrated negative geometries have also been studied for four points in the ABJM theory \cite{He:2022cup,Henn:2023pkc,Lagares:2024epo,Li:2024lbw}, and it would be interesting to find an application of the results from this paper also in this case.

\section{Acknowledgements}
We thank the Galileo Galilei Institute for Theoretical Physics for the hospitality and the INFN for partial support during the completion of this work. RG would like to thank Michele Santagata for interesting discussions on related topics and Jaroslav Trnka for correspondence on extension of these results beyond ladder topologies.

\bibliographystyle{nb}

\bibliography{ladders_in_loops_bib}

\begin{thebibliography}{10}
\ifx\href\asklfhas\newcommand{\href}[2]{#2}\fi
\ifx\arxivref\asklfhas\newcommand{\arxivref}[2]{\href{http://arxiv.org/abs/#1}{#2}}\fi
\ifx\doiref\asklfhas\newcommand{\doiref}[2]{\href{http://dx.doi.org/#1}{#2}}\fi
\raggedright
\small
\parskip 0pt

\bibitem{Arkani-Hamed:2017tmz}
N.~Arkani-Hamed, Y.~Bai and T.~Lam,
\textit{``{Positive Geometries and Canonical Forms}''},
\textsf{\doiref{10.1007/JHEP11(2017)039}{JHEP~1711,~039~(2017)}},
\texttt{\arxivref{1703.04541}{arxiv:1703.04541}}.

\bibitem{Arkani-Hamed:2013jha}
N.~Arkani-Hamed and J.~Trnka,
\textit{``{The Amplituhedron}''},
\textsf{\doiref{10.1007/JHEP10(2014)030}{JHEP~1410,~030~(2014)}},
\texttt{\arxivref{1312.2007}{arxiv:1312.2007}}.

\bibitem{He:2022cup}
S.~He, C.-K.~Kuo, Z.~Li and Y.-Q.~Zhang,
\textit{``{All-Loop Four-Point Aharony-Bergman-Jafferis-Maldacena Amplitudes
  from Dimensional Reduction of the Amplituhedron}''},
\textsf{\doiref{10.1103/PhysRevLett.129.221604}{Phys.~Rev.~Lett.~129,~221604~(2022)}},
\texttt{\arxivref{2204.08297}{arxiv:2204.08297}}.

\bibitem{He:2023rou}
S.~He, Y.-t.~Huang and C.-K.~Kuo,
\textit{``{The ABJM Amplituhedron}''},
\textsf{\doiref{10.1007/JHEP09(2023)165}{JHEP~2309,~165~(2023)}},
\texttt{\arxivref{2306.00951}{arxiv:2306.00951}},
[Erratum: JHEP 04, 064 (2024)].

\bibitem{Arkani-Hamed:2017mur}
N.~Arkani-Hamed, Y.~Bai, S.~He and G.~Yan,
\textit{``{Scattering Forms and the Positive Geometry of Kinematics, Color and
  the Worldsheet}''},
\textsf{\doiref{10.1007/JHEP05(2018)096}{JHEP~1805,~096~(2018)}},
\texttt{\arxivref{1711.09102}{arxiv:1711.09102}}.

\bibitem{Eden:2017fow}
B.~Eden, P.~Heslop and L.~Mason,
\textit{``{The Correlahedron}''},
\textsf{\doiref{10.1007/JHEP09(2017)156}{JHEP~1709,~156~(2017)}},
\texttt{\arxivref{1701.00453}{arxiv:1701.00453}}.

\bibitem{He:2024xed}
S.~He, Y.-t.~Huang and C.-K.~Kuo,
\textit{``{All-loop geometry for four-point correlation functions}''},
\textsf{\doiref{10.1103/PhysRevD.110.L081701}{Phys.~Rev.~D~110,~L081701~(2024)}},
\texttt{\arxivref{2405.20292}{arxiv:2405.20292}}.

\bibitem{Damgaard:2019ztj}
D.~Damgaard, L.~Ferro, T.~Lukowski and M.~Parisi,
\textit{``{The Momentum Amplituhedron}''},
\textsf{\doiref{10.1007/JHEP08(2019)042}{JHEP~1908,~042~(2019)}},
\texttt{\arxivref{1905.04216}{arxiv:1905.04216}}.

\bibitem{Ferro:2022abq}
L.~Ferro and T.~Lukowski,
\textit{``{The Loop Momentum Amplituhedron}''},
\textsf{\doiref{10.1007/JHEP05(2023)183}{JHEP~2305,~183~(2023)}},
\texttt{\arxivref{2210.01127}{arxiv:2210.01127}}.

\bibitem{Ferro:2023qdp}
L.~Ferro, R.~Glew, T.~Lukowski and J.~Stalknecht,
\textit{``{Prescriptive unitarity from positive geometries}''},
\textsf{\doiref{10.1007/JHEP03(2024)001}{JHEP~2403,~001~(2024)}},
\texttt{\arxivref{2308.02438}{arxiv:2308.02438}}.

\bibitem{Lukowski:2023nnf}
T.~Lukowski and J.~Stalknecht,
\textit{``{Momentum Amplituhedron for N=6 Chern-Simons-Matter Theory:
  Scattering Amplitudes from Configurations of Points in Minkowski Space}''},
\textsf{\doiref{10.1103/PhysRevLett.131.161601}{Phys.~Rev.~Lett.~131,~161601~(2023)}},
\texttt{\arxivref{2306.07312}{arxiv:2306.07312}}.

\bibitem{Ferro:2024vwn}
L.~Ferro, R.~Glew, T.~Lukowski and J.~Stalknecht,
\textit{``{The Two-loop MHV Momentum Amplituhedron from Fibrations of
  Fibrations}''},
\texttt{\arxivref{2407.12906}{arxiv:2407.12906}}.

\bibitem{Parisi:2021oql}
M.~Parisi, M.~Sherman-Bennett and L.~K.~Williams,
\textit{``{The m=2 amplituhedron and the hypersimplex: signs, clusters,
  tilings, Eulerian numbers.}''},
\textsf{\doiref{10.1090/cams/23}{Commun.~Am.~Math.~Soc.~3,~329~(2023)}},
\texttt{\arxivref{2104.08254}{arxiv:2104.08254}}.

\bibitem{Arkani-Hamed:2021iya}
N.~Arkani-Hamed, J.~Henn and J.~Trnka,
\textit{``{Nonperturbative negative geometries: amplitudes at strong coupling
  and the amplituhedron}''},
\textsf{\doiref{10.1007/JHEP03(2022)108}{JHEP~2203,~108~(2022)}},
\texttt{\arxivref{2112.06956}{arxiv:2112.06956}}.

\bibitem{Brown:2023mqi}
T.~V.~Brown, U.~Oktem, S.~Paranjape and J.~Trnka,
\textit{``{Loops of loops expansion in the amplituhedron}''},
\textsf{\doiref{10.1007/JHEP07(2024)025}{JHEP~2407,~025~(2024)}},
\texttt{\arxivref{2312.17736}{arxiv:2312.17736}}.

\bibitem{Chicherin:2024hes}
D.~Chicherin, J.~Henn, J.~Trnka and S.-Q.~Zhang,
\textit{``{Positivity properties of five-point two-loop Wilson loops with
  Lagrangian insertion}''},
\texttt{\arxivref{2410.11456}{arxiv:2410.11456}}.

\bibitem{Arkani-Hamed:2010zjl}
N.~Arkani-Hamed, J.~L.~Bourjaily, F.~Cachazo, S.~Caron-Huot and J.~Trnka,
\textit{``{The All-Loop Integrand For Scattering Amplitudes in Planar N=4
  SYM}''},
\textsf{\doiref{10.1007/JHEP01(2011)041}{JHEP~1101,~041~(2011)}},
\texttt{\arxivref{1008.2958}{arxiv:1008.2958}}.

\bibitem{Arkani-Hamed:2010pyv}
N.~Arkani-Hamed, J.~L.~Bourjaily, F.~Cachazo and J.~Trnka,
\textit{``{Local Integrals for Planar Scattering Amplitudes}''},
\textsf{\doiref{10.1007/JHEP06(2012)125}{JHEP~1206,~125~(2012)}},
\texttt{\arxivref{1012.6032}{arxiv:1012.6032}}.

\bibitem{Henn:2023pkc}
J.~M.~Henn, M.~Lagares and S.-Q.~Zhang,
\textit{``{Integrated negative geometries in ABJM}''},
\textsf{\doiref{10.1007/JHEP05(2023)112}{JHEP~2305,~112~(2023)}},
\texttt{\arxivref{2303.02996}{arxiv:2303.02996}}.

\bibitem{Lagares:2024epo}
M.~Lagares and S.-Q.~Zhang,
\textit{``{Higher-loop integrated negative geometries in ABJM}''},
\textsf{\doiref{10.1007/JHEP05(2024)142}{JHEP~2405,~142~(2024)}},
\texttt{\arxivref{2402.17432}{arxiv:2402.17432}}.

\bibitem{Li:2024lbw}
Z.~Li,
\textit{``{Integrating the full four-loop negative geometries and all-loop
  ladder-type negative geometries in ABJM theory}''},
\textsf{\doiref{10.1007/JHEP10(2024)124}{JHEP~2410,~124~(2024)}},
\texttt{\arxivref{2402.17023}{arxiv:2402.17023}}.

\end{thebibliography}

\end{document}